\documentclass[pre,aps,amsfonts,amssymb,multicol,epsfig,graphics,
referee]{revtex4}

\usepackage{graphicx}
\usepackage{bm}

\def \beq{\begin{equation}}         \def \eeq{\end{equation}}
\def \beqa{\begin{eqnarray}}        \def \eeqa{\end{eqnarray}}
\def \bea{\begin{array}}        \def \eea{\end{array}}

\def\bio#1#2#3{{Biophys. J. }{\bf #1}, #2 (#3)}

\def\jmb#1#2#3{{J. Mol. Biol. }{\bf #1}, #2 (#3)}

\def\jpc#1#2#3{{J. Phys. Chem. B }{\bf #1}, #2 (#3)}
 
\def\jcp#1#2#3{{J. Chem. Phys. }{\bf #1}, #2 (#3)}

\def\nat#1#2#3{{Nature(London) }{\bf #1}, #2 (#3)}

\def\nats#1#2#3{{Nature struct. Biol. }{\bf #1}, #2 (#3)}
\def\pnas#1#2#3{{Proc. Natl. Acad. Sci. USA }{\bf #1}, #2 (#3)}
\def\pre#1#2#3{{Phys. Rev. E }{\bf #1}, #2 (#3)}

\def\prl#1#2#3{{Phys. Rev. Lett. }{\bf #1}, #2 (#3)}
\def\sci#1#2#3{{Science }{\bf #1}, #2 (#3)}

\usepackage{amsmath}
\usepackage{epsfig}

\begin{document}

\title{A model for the force stretching double-stranded chain molecules}
\author{Fei Liu}
\email[]{liufei@itp.ac.cn}
\affiliation{Institute of Theoretical Physics, The Chinese 
Academy of Sciences, P. O. Box 2735, Beijing 100080, China}
\author{Luru Dai}
\affiliation{Institute of Theoretical Physics, The Chinese 
Academy of Sciences, P. O. Box 2735, Beijing 100080, China}
\author{Zhong-can Ou-Yang}
\affiliation{Institute of Theoretical Physics, The Chinese 
Academy of Sciences, P. O. Box 2735, Beijing 100080, China}
\affiliation{Center for Advanced Study, Tsinghua University, Beijing 100084, 
China}

\date{\today}

\begin{abstract}
We modify and extend the recently developed statistical mechanical model for 
predicting the thermodynamic properties of chain molecules having noncovalent 
double-stranded conformations, as in RNA or ssDNA, and $\beta-$sheets in 
protein, by including the constant force stretching at 
one end of molecules as in a typical single-molecule experiment. The 
conformations of double-stranded regions of the chain are calculated based on 
polymer graph-theoretic approach [S-J. Chen and K. A. Dill, J. Chem. Phys. 
{\bf109}, 4602(1998)], while the unpaired single-stranded regions are treated 
as self-avoiding walks. Sequence dependence and excluded volume interaction 
are taken into account explicitly. Two classes of conformations, hairpin and 
RNA secondary structure are explored. For the hairpin conformations, all possible 
end-to-end distances corresponding to the different types of double-stranded 
regions are enumerated exactly. For the RNA secondary structure 
conformations, a new recursive formula incorporating the secondary structure 
and end-to-end distribution has been derived. Using the model, we 
investigate the extension-force curves, contact and population distributions 
and re-entering phenomena, respectively. we find that the force stretching 
homogeneous chains of hairpin and secondary structure conformations are very 
different: the unfolding of hairpins is two-state, while unfolding the latter 
is one-state. In addition, re-entering transitions only present in hairpin 
conformations, but are not observed in secondary structure conformations. 
\end{abstract}
\maketitle

\section{Introduction}
\label{sec-intro}
In recent years, diverse mechanical manipulation techniques of 
single-molecule,  such as optical tweezers, atomic force microscopy, and 
soft microneedles have been applied in probing basic mechanical, physical 
and chemical properties of single biological macromolecule, such as proteins, 
nucleic acids, and molecular motors. The basic principle is to exert force 
in the $pN$ range on the studied molecules, and then force-extension (FECs) 
or extension-force curves (EFCs) are recorded\cite{smith,kishino,riefp,visscher}. 
Many useful insights about biomolecules were obtained by analyzing these 
curves, going from the detailed elastic properties\cite{smith} to 
complex structure 

experiments\cite{essevaz,bockelmann,maier,rief,liphardt}. In experiments of 
unzipping of double-stranded DNA (dsDNA), FECs show gross features, which  
reflect simply the local GC versus AT content along the sequence\cite{essevaz,
bockelmann}. While in stretching single-stranded DNA (ssDNA), very 
different EFCs have been observed: the EFCs recorded by Maier 
{\it{et al.}} showed extremely smooth\cite{maier}; in contrast, Rief 
{\it{et al.}} found that ssDNA end-to-end distance (EED) suddenly 
changed during narrow force range. In recent unzipping shorter RNA 
experiment, except EED increasing abruptly, EFCs with intermediates 
were observed\cite{liphardt}. 

On the theoretical side, compared to the great effort in understanding 
the unzipping of dsDNA\cite{thompso,cocco,lubensky,marenduzzo}, little 
attention has been paid to the RNA or ssDNA stretching 
problem\cite{montanari, muller225,zhou1,zhou2,zhang,muller348,gerland}. 
Montanari and M$\acute{e}$zard discussed the EFCs of a homogeneous 
ssDNA sequence. Their model exhibited a second order phase transition 
from a collapsed globular state to an extended necklace-like 
phase\cite{montanari}. Zhou and Zhang pointed out the important roles 
of electrostatic and base-pair stacking interactions played in 
stretching ssDNA\cite{zhou2,zhang}. M$\ddot{u}$ller {\it{et. al.}} 
considered how the random disorder sequence 
modifies EFCs characteristics\cite{muller348}. Gerland {\it et al.} 
explored quantitatively how nuclei acids structure determines a broad range 
of FECs from very rugged to completely featureless. They found that apart 
from the entropy elasticity of the unpaired single strand\cite{thompso}, 
there are two additional mechanism, the compensation effect and the 
contribution of suboptimal secondary structures smoothing the features of 
FECs\cite{gerland}. Though their results were discussed on the constant 
extension ensemble, it may be true for constant force 
ensemble\cite{lubensky}. 

There are several reasons to attract increasing attention on the force 
stretching RNA or ssDNA. First, studies of stretching RNA may generate new 
insights into the RNA folding problem\cite{tinoco}, including folding 
pathways and the energy landscape\cite{chen00}. Unlike the simpler 
hairpin structure, such as dsDNA, RNA may fold into complicated 
branched structures, and the native state has to compete with a large number 
of suboptimal states. Exerting force on the molecule will drive the native 
state into metastable states, which might be seen from the FECs or 
EFCs. Second, force stretching provides a direct approach to measure RNA 
thermodynamic parameters, which are key in predicting RNA secondary 
structure\cite{tinoco,liphardt}. In addition, the force stretching RNA is 
also an intriguing physical problem, e.g., how RNA secondary structures 
determine the outcoming EFCs\cite{gerland}, and how to describe the RNA 
properties far from equilibrium induced by external force is still an open 
question\cite{lubensky}. Because the structure complexity of RNA is 
intermediate between dsDNA and protein, the exploration may also be valuable in 
understanding properties of dsDNA and protein\cite{riefp}. 

Although the theoretical models about force stretching RNA 
molecules\cite{montanari,zhou1,zhang,gerland} have been proposed, more 
realistic and detailed models are needed\cite{tinoco}. We note that 
with a few exceptions\cite{gerland}, almost all models were restricted the 
study of homogeneous or completely random RNA sequences. Whereas stretching 
experiments showed that sequences affect EFCs dramatically\cite{maier,
rief,liphardt}. In very recent 
experiment\cite{dessinges}, Dessinges {\it et. al } found that 
excluded-volume interactions play an essential role as stretching 
ssDNA, which were assumed to be negligible before. To take into account 
the two factors, in this paper we will modify and 
extend the statistical model of double-stranded chain conformations 
developed by Chen and Dill to constant force stretching 
ensemble\cite{chen95,chen98}. We think that their model is suitable 
for our aim. First, the model retains a relatively high degree of 
realism, which the sequence dependence, 
nonlocal intrachain contacts, and the excluded volume interactions are 
accounted for explicitly. Second, the model is more ``microscopic", i.e., 
only one parameter, base-pair contact energy is required. The entropy is 
calculated from the number of conformations directly. In addition, 
because the model can completely characterize the full energy landscape 
of secondary structure having chain lengths up to hundreds 
bases\cite{chen00}, the exploration of how force changes the folding pathways or 
energy landscape of RNA is valuable. 

The organization of the paper is as follows. We first, in Sec~
\ref{reviewpolymer}, simply overview the statistical model of 
double-stranded chain molecules developed by Chen and Dill. 
In Sec.~\ref{stretchdouble}, we extend the models about two classes of 
chain molecules, hairpin and RNA-like 
secondary structures conformations to constant force stretching 
cases, respectively. 
Their corresponding EFCs, contact probability, 
population distribution and re-entering phenomena, and the difference 
are discussed in Sec.~\ref{disscussion}. Section.~\ref{conclusion} is our 
conclusion. Some numerical process are relegated to 
appendices. In the Appendix~\ref{endtoendhairpin} we illustrate the 
calculation of EED with a special case in hairpin conformations, namely 
when the closed graph is of type 1. In the 
Appendix~\ref{extrapolate}, we simply describe how to get the 
EED distribution of ``open" self-avoiding walk (OSAW) (self-avoiding 
walks involving no neighboring contacts) along one coordinate through 
extrapolating approach.

\section{The double-stranded chain model}
\label{reviewpolymer}
The details of the statistical model of double-stranded chain molecules 
can be found in Refs. \cite{chen95} and \cite{chen98}. Here we give a brief 
review.
\subsection{Polymer graph theory}
\begin{figure}[htpb]
\begin{center}
\includegraphics[width=0.5\columnwidth]{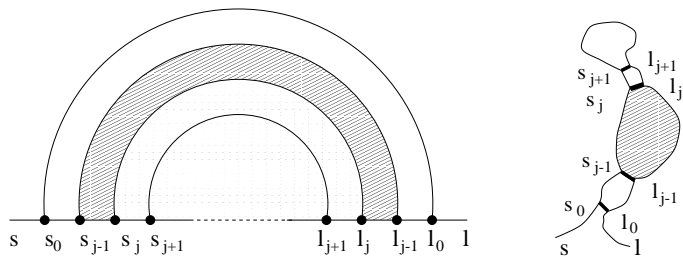}
\caption{The polymer graph of hairpin conformations. The shadowed region is 
a face of the graph. The graph is divided into two parts: tow tails 
$[s,s_0]$ and $[l_0,l]$, and one double-stranded region, or CG  
enclosed by outmost link $(s_0,l_0)$. Here $s$ and $l$ are two end monomers in 
chain. The double-stranded region is composed of links nested each other: 
$(s_0,l_0),\cdots(s_{j-1},l_{j-1}),(s_j,l_j), (s_{j+1},l_{j+1})$.}
\label{polygraph}
\end{center}
\end{figure}

The model is based on polymer graphs, diagrammatic representations of the 
self-contacts made by different chain conformations. Fig.~\ref{polygraph} 
shows a hairpin conformation and corresponding polymer graph: vertices 
represent the chain monomers, straight line links symbolize the covalent 
bonds, and curved links stand for spatial contact between monomers.
A given polymer graph represents an ensemble of chain conformations that are 
consistent with given contacts. Conformations having contacts other 
than those specified by the polymer graph belong to other graphs. 
In general, fewer curved links exist in a graph, a more larger number of 
chain conformations are consist with this graph. Any two pairs of curved 
links in a polymer graph must bear one of three relationships: nested, 
unrelated and crossing linked. 
Graphs of the double-stranded chain conformations involve no crossing 
links, examples of which include the simplest hairpin structure, such 
as dsDNA, and mainly secondary structures among nucleic acids and antiparallel 
$\beta$-sheets in proteins. In terms of polymer graph, the partition function 
of a $(N+1)$ monomers ((N+1)-mer) chain molecule is given as a sum over all 
possible polymer graphs, 
\begin{eqnarray}
\label{graphsum}
{\cal Q}_N(T)=\sum\limits_{\Gamma}\Omega(\Gamma)e^{-E(\Gamma)/k_B T},
\end{eqnarray}
where $k_B$ is Boltzmann constant, $T$ is temperature, $\Gamma$ is an index of a possible polymer graph, $E(\Gamma)$ and $\Omega(\Gamma)$ are the energy 
and the number of conformations of the given polymer graph $\Gamma$, 
respectively. To calculate $\Omega(\Gamma)$, 
Chen and Dill developed a matrix multiplication method\cite{chen95,chen98}.

\subsection{The matrix method for a given polymer graph}
\begin{figure}[htpb]
\begin{center}
\includegraphics[width=0.7\columnwidth]{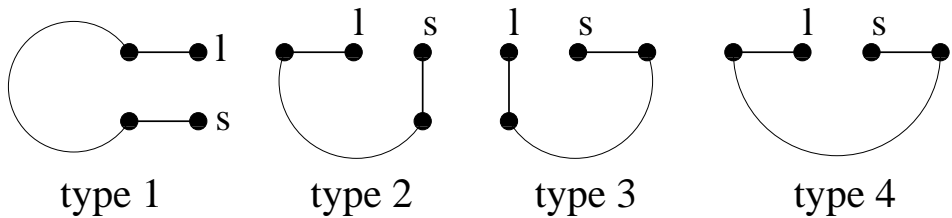}
\caption{ The four types of port (inlet or outlet) shapes on 2D lattice.} 
\label{portshap}
\end{center}
\end{figure}

A complex polymer graph can be divided into a 
series of faces consecutively, in which  
face is a region in graph that is bounded by curved and straight lines 
and contains no other edges; see the shadowed region in Fig.~\ref{polygraph}. 
Faces are classified into five types: left (L), middle (M), right (R), 
left-right (LR) and isolated (I), according to the arrangement of the nested 
curved links that bound the face. The calculation of the full partition 
function $\Omega(\Gamma)$ for a given  graph $\Gamma$ is correspondingly 
separated into two steps: to count all conformations for each face, as if 
they are isolated and independent of each other, then to assemble these  
conformations into $\Omega(\Gamma)$. To avoid that conformations of different 
faces bump into each other (excluded-volume interaction), more detailed 
information about the conformations of faces is needed. On two dimension (2D) 
lattice, it is realized by exact enumeration. First, conformations of each 
face are classified into sixteen types according to the shape of ports 
(inlet and outlet) through which it is connected to other faces; see 
Fig.~\ref{portshap}. Then the compatibility between neighboring faces 
can be checked exactly through the spatial compatibility between the outlet 
of one face and the inlet of next face. It is convenient to 
introduce two matrices: the Face Count Matrix (FCM) ${\bf S}_t$, 
which matrix element $({\bf S}_t)_{ij}$ is the number of conformations having 
an inlet conformation of type $i$ ($1\le i\le 4$) and an outlet 
conformation of type $j$ ($1\le j\le 4$) for a type $t$ face, and 
viability matrix ${\bf Y}_{t_1t_2}$, which element 
$({{\bf Y}_{t_1t_2}})_{ij}$ is $1$ or $0$ if connection 
between type $i$ outlet of a type $t_1$ face and type $j$ inlet of a type 
$t_2$ face is viable or not viable.  For a hairpin graph $\Gamma$ having 
$M$ faces, $\Omega(\Gamma)$ is derived as a product of matrices:  
\begin{eqnarray}
\label{omegcalculate}
\Omega(\Gamma)={\bf U}\cdot{\bf S}_{t_M}\cdot{\bf Y_{t_Mt_{M-1}}}\cdot{\bf S}_{t_{M-1}}\cdots{{\bf S}_{t_1}}\cdot{\bf U}^t,
\end{eqnarray}
where ${\bf U}=\{1,1,1,1\}$, ${\bf U}^{t}$ is the transpose of 
${\bf U}$. 

\subsection{The partition function}
In order to calculate the partition function of a whole chain using 
Eq.~\ref{graphsum}, the sum over all possible polymer graphs is 
necessary. For the double-stranded conformations, an efficient dynamic 
programming algorithm was developed in Ref. ~\cite{chen98}. 
The idea is to start with a short chain segment and elongate the segment by 
adding one monomer for each step, and then to calculate recursively the partition 
function of the longer segment using the result of the shorter one. 
The algorithms will be discussed as needed in following sections. More 
useful alternative expression of Eq.~\ref{graphsum} is 
\begin{eqnarray}
\label{engsum}
{\cal Q}_N(T)=\sum\limits_E g_N(E)e^{-E/k_B T},
\end{eqnarray}
where $g_N(E)$ is the density of states, or the number of conformations having  
energy $E$, which is defined as 
\begin{eqnarray}
g_N(E)=\sum\limits_{\Gamma}\left.\Omega(\Gamma)\right|_{E(\Gamma)=E}.
\end{eqnarray}

\section{Force stretching double-stranded chain molecule: the hairpin 
and RNA secondary structure conformations}
\label{stretchdouble}
\begin{figure}[htpb]
\begin{center}
\includegraphics[width=0.8\columnwidth]{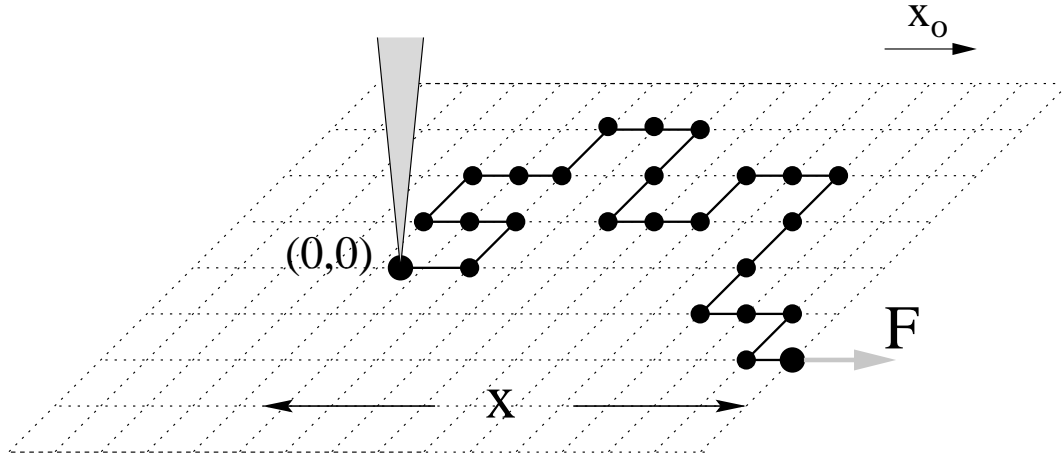}
\caption{Sketch of the force stretching experiment on the 2D lattice. The 
larger dark points represent two ends of a chain molecule. One end 
is fixed by a pin, and another is stretched by a constant 
force ${\bf F}=f{\bf x}_o$ along a given unit vector ${\bf x}_o$. The 
monomers are denoted to be small dark points. The average extension $x$ is  
recorded as a function of $f$.}
\label{apparatus}
\end{center}
\end{figure}

Fig.~\ref{apparatus} depicts the situation studied in this paper: one end 
of double-stranded chain molecule is fixed by a pin, and a constant force 
${\bf F}=f{\bf x}_o$ is exerted on the other end, where ${\bf x}_o$ is a given 
unit vector; average extension $x$ along ${\bf x}_o$ is recorded as 
function of $f$. To include the contribution of force stretching energy, we 
extend the partition function Eq.~\ref{engsum} to 
${\cal Q}_N(T;f)$ at given force $f$ as 
\begin{eqnarray}
\label{engsumforc}
{\cal Q}_N(T;f)=\sum\limits_E\sum\limits_\Delta g_N(E;\Delta)e^{
-\beta(E-f\Delta)},
\end{eqnarray}
where $\Delta$ is EED of the chain along ${\bf x}_o$, and $g_N(E;\Delta)$ is 
the number of conformations whose energy and EED are $E$ and $\Delta$, 
respectively. Define an auxiliary partition function 
$\mathcal{G}_N(E;f)$ as 
\begin{eqnarray}
\label{eedsum}
\mathcal{G}_N(E;f)=\sum\limits_\Delta g_N(E;\Delta)e^{\beta{f\Delta}}.
\end{eqnarray}
Then instead of Eq.~\ref{engsumforc}, we have 
\begin{eqnarray}
\label{newengsum}
{\cal Q}_N(T;f)=\sum\limits_E\mathcal{G}_N(E;f)e^{-\beta E}.
\end{eqnarray}
Apparently, the following relations are satisfied: 
$\sum_\Delta g_N(E;\Delta)=g_N(E)$ or $\mathcal{G}_N(E;f=0)=g_N(E)$. The 
measured extension $x$ at force $f$ is calculated from partition 
function as  
\begin{eqnarray}
\label{extenforce}
x=k_BT\frac{\partial}{\partial f}\log{\cal Q}_N(T;f).
\end{eqnarray}

The calculation of $g_N(E;\Delta)$ is key in force stretching problem. Due to   
the new parameter, EED $\Delta$, the dynamic programming 
algorithm\cite{chen98} must be modified and extended carefully. In following 
sections, we show how to compute $g_N(E;\Delta)$ of 
hairpin and RNA secondary structure conformations respectively. 

\subsection{Force stretching hairpin conformations}\label{stretchhp} 
As one of the simplest elements in secondary structure, hairpin 
exists in a large class of biomolecules, such as RNA hairpin, peptide 
$\beta$-hairpin and dsDNA hairpin. Recent works showed that the 
hairpin conformations have remarkable thermodynamic and kinetic 
behaviors\cite{munoz,chen98}. One may hope that they also have remarkable 
mechanical characteristic. On the other hand, more precise formula and 
the essence of theory for complex RNA secondary structure conformations 
make us to explore them stretched by force independently. 

The polymer graphs of hairpin conformations are that every curved 
link bears a nested relationship with respect to every other curved link.
The polymer graph is composed of two parts: two non-self-contacting 
tail chains $(s,s_0)$ and $(l_0,l)$ and one double-stranded region 
$(s_0,l_0)$, which is defined as Closed Graph (CG)\cite{chen98}; 
see Fig.~\ref{polygraph}. The number of conformations for a given graph 
without force equals a multiplication of the number of double-stranded 
conformations and 
the number of two tails conformations. In terms of four types of the 
outermost faces, the graphs are classified into four types. (LR type is 
excluded from hairpin conformations). To sum over all polymer graphs, two 
matrices, the Closed Graph Count Matrix 
(CGCM), ${{\bf G^\ast}_t}\left[E,s_0,l_0\right]$ and diagonal matrix 
${\bf \omega}\left[s,s_0;l_0,l\right]$ have been defined: 
$({{\bf G^\ast}_t}\left[E,s_0,l_0\right])_{ij}$ is the sum over the number 
of conformations for all possible $t$ type graphs having energy $E$, 
given that the outmost link spans from vertex $l_0$ to vertex $s_0$, and 
innermost and outermost links are in $i$ and $j$ types conformations; 
diagonal matrix element $({\bf \omega}\left[s,s_0;l_0,l\right])_{ii}$ is the 
number of conformations of two tails $(s,s_0)$ and $(l_0,l)$ that are 
spatially compatible with type $i$ conformations. State density $g_N(E)$ then 
can be written as: 
\begin{eqnarray}
g_N(E)={\bf U}\cdot\sum_{s_0, l_0, t}{\bf\omega}\left[s,s_0;l_0,l\right]
\cdot{\bf G^\ast}_t\left[E,s_0,l_0\right] \cdot{\bf U}^t,
\label{hairpinsum}
\end{eqnarray}
here $1\le t\le 4$\cite{chen98}.

\begin{figure}[htpb]
\begin{center}
\includegraphics[width=1.\columnwidth]{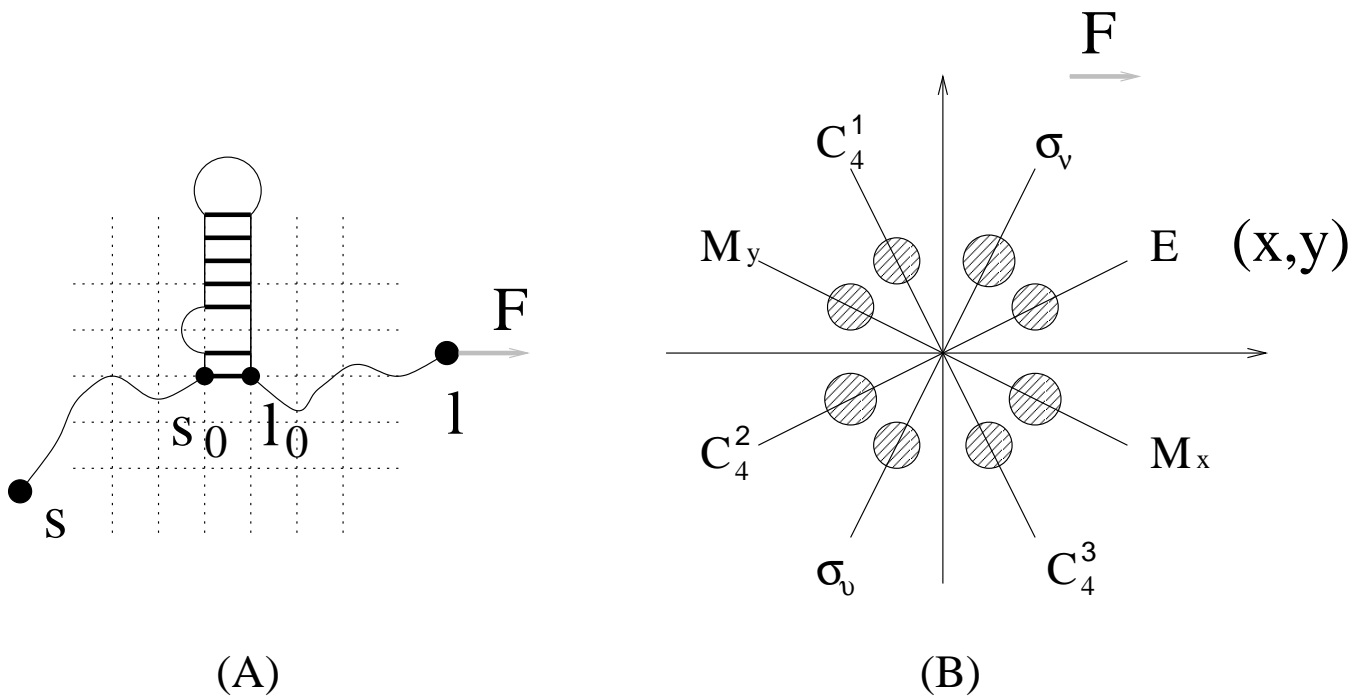}
\caption{(A) Sketch of the force stretching hairpin conformations. Here the 
double-stranded region contributes $+1$ to EED of the whole 
chain. (B) Illustration of how the conformation ended at $(x,y)$ is  
distributed to whole lattice plane by eight square symmetric transformations, 
where the shadowed circles represent the double-stranded region. 
}
\label{hairpinforc}
\end{center}
\end{figure}

When force is nonzero, $g_N(E)$ is extended to $g_N(E;\Delta)$. ${\bf \omega}[s,s_0;l_0,l]$ is also {\it{nondegenerated}} into 
${\bf \omega}[s,s_0;l_0,l|\Delta]$. From Eq.~\ref{hairpinsum}, we note that 
force stretching work just depends on EED of the chain, which does not involve 
any detailed double-stranded structure directly, i.e., only the outmost link 
$(s_0,l_0)$ contributes $\pm 1$ or $0$ to EED $\Delta$ of the chain along 
${\bf x}_0$ direction; see Fig.~\ref{hairpinforc}. Though, in real nucleic 
acid, a typical distance of a hydrogen bond is about $3$ times than 
the nucleotide distance, it does not make sense for our description on a 
coarse-grained level. We can exactly enumerate EED of 
the whole chain with three steps: first, to fix one conformation of a given 
CG on lattice and grow two tails which are compatible 
with the graph type; then, to enumerate EED of the two tails and combine 
them with EED of the graph according to force direction; and finally, to 
distribute the full conformation to lattice plane by eight square symmetry 
transformations ($D_4$ group). We define a new diagonal matrix 
$\Pi\left[s,s_0;l_0,l|f\right]$,
\begin{eqnarray}
\Pi\left[s,s_0;l_0,l|f\right]=
\sum\limits_\Delta{\bf\omega}\left[s,s_0;l_0,l|\Delta\right]
e^{\beta{f\Delta}}.
\end{eqnarray}
So we obtain      
\begin{eqnarray}
\mathcal{G}_N(E;f)=
{\bf U}\cdot\sum_{s_0, l_0, t} \Pi\left[s,s_0;l_0,l|f\right] 
\cdot{\bf G^\ast}_t \left[E,s_0,l_0\right] \cdot{\bf U}^t.
\label{hairpinused}
\end{eqnarray}

The basic idea is very simple, however, it is very cumbersome to complete 
this process. Rather than to give a detailed analysis here, we 
illustrate the process in Appendix~\ref{endtoendhairpin} by a special case. 

Though the enumeration method gives quite accurate EED of the hairpin 
conformations, unfortunately, as applied to longer molecule chain, 
it is hard to give exactly analytical formula about $n_x^t(N;x)$ and 
$n_y^t(N;y)$,  which are defined in Appendix~\ref{endtoendhairpin}. A more 
practical approximation is proposed in next section. 

\subsection{Force stretching secondary structure conformations}
\label{stretchrna}
Secondary structure can be seen as a tree-like structure into which four 
basic structure elements, helix, loops, bulges and junctions compose 
through self-similarity arrangement.  Each graph of secondary structure 
conformations is divided into two levels: the first lever is a combination 
of unrelated double-stranded regions connected with single-stranded chains; 
the second level is that each CG may be viewed as an independent 
secondary structure except that two end monomers of the region contact.  
We first simply review how to calculate the state density $g(E)$ without 
force. Necessary definitions are introduced. 

To be different from hairpin conformations, the calculation of full state 
density of secondary structure is more complex. First all possible CGCMs, 
${{\bf G^\ast}_t}\left[E,a,b\right]$ (Note that LR-type faces are included) 
are computed, where $(a,b)$ is the outmost link of 
the CG. Any CG is composed of smaller 
unrelated subclosed graphs, auxiliary matrices, ${\bf K}_t[E,l,a,b]$ counting 
a combination of conformations of all subgraphs are introduced, where 
${\it cycle length }$ $l$ is total number of monomers of the 
single-stranded chains in $[a,b]$\cite{chen98}. CGCMs can be obtained by 
multiplication of matrix ${\bf K}_t[E,l,a,b]$ and the number of 
conformations of the single-stranded part. Then in order to combine 
conformations of unrelated graphs into whole, 
matrices ${\bf G}_t\left[E,s,a\right]$ having energy $E$ for full polymer 
graph $[s,a]$ are defined, where $t=0,1,2$ represent three full graph 
types which are classified depending on whether $0$, $1$ and $2$ existing 
links are connected to the rightmost monomer, see Fig.~\ref{recurse}. Their 
element $({\bf G}_t\left[E,s,a\right])_{ij}$ is the number of conformations for 
graphs in which the outmost link (of rightmost subgraph) and the 
innermost link (of the leftmost subgraph) are in $j$ and $i$ conformations, 
respectively. All matrices mentioned above are calculated by 
dynamics programming algorithm\cite{chen98}. The full density of states of 
the secondary structure conformations is written as
\cite{chen98} 
\begin{eqnarray}
g_N(E)={\bf U}\cdot\sum_{t=0}^{2}{\bf G}_t\left[E,s,l\right] \cdot{\bf U}
^t.
\label{secstrusum}
\end{eqnarray}

When force is exerted on the chain, exact knowledge about EED  
is required. In principle, we can generalize the enumeration method 
in hairpin case to secondary structure: to enumerate all 
EEDs of each single-stranded chains separated by CGs, and to 
combine them with contribution of the CGs, $\pm1$ and $0$ to get the full 
EED. Apart from several special cases, e.g., only one or two CGs 
allowed in polymer graph, however, the approach is not practicable 
for general graphs from the experience in counting EEDs of hairpin conformations.  
We propose an alternative method. Because that the detailed structures of 
the CGs are not involved into the whole EED directly, CGs can be viewed as 
{\it effective} covalent bonds connecting left and right parts 
of conformations. Hence, the secondary structure conformations are 
identified into OSAWs with the reduced monomers; the effective 
bonds bear the excluded volume interaction of CGs with other units in 
the chain. The EEDs are computed by enumeration and extrapolation 
method. Numerical results of the latter can be found in 
Appendix~\ref{extrapolate}. Considering that conformations of monomers 
neighboring CGs are almost ``frozen" by excluded volume interaction, 
it may be more reasonable to include more monomers into the effective 
bonds such as illustrated in Fig.~\ref{renormchain}. In present paper the 
simplest case is discussed. Though the effective chain approach (ECA) may 
overestimate the number of conformations for partially considering 
excluded volume interaction, it is still valuable before better methods 
are given.

To realize the ECA, we modify the matrices ${\bf G}_t\left[E,s,a\right]$ to 
${\bf G}_t\left[E,n,s,a\right]$, where new parameter $n$ is  
the number of effective chain monomers. The recursive relations in 
Ref.~\cite{chen98} (also see Fig.~\ref{recurse}) then are extended into following 
relations:

\begin{figure}
\begin{center}
\includegraphics[width=0.6\columnwidth]{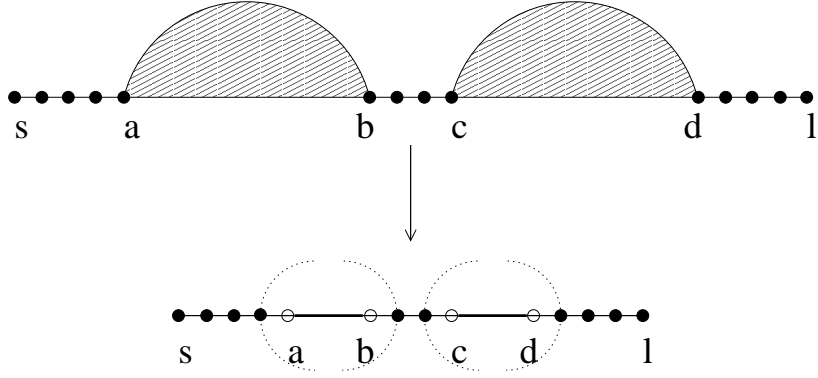}
\caption{ Illustration of how unrelated closed polymer graphs are reduced into 
effective chain to calculate EED of the chain. For example, two CGs $[a,b]$ and $[c,d]$ in upper are replaced by two bonds in below, and 
length of the chain is reduced to $(a-s)+(c-b)+(l-d)+2$ from $l-s$. 
Considering that excluded volume interactions between CGs and single-strand 
parts make conformations of the neighboring monomers of CGs to be ``frozen", 
it may be reasonable to reduce chain length more, showing here by dashed 
brackets, and effective length is $(a-s)+(c-b)+(l-d)-2$ now.}
\label{renormchain}
\end{center}
\end{figure}

\begin{figure}[htpb]
\begin{center}
\includegraphics[width=0.8\columnwidth]{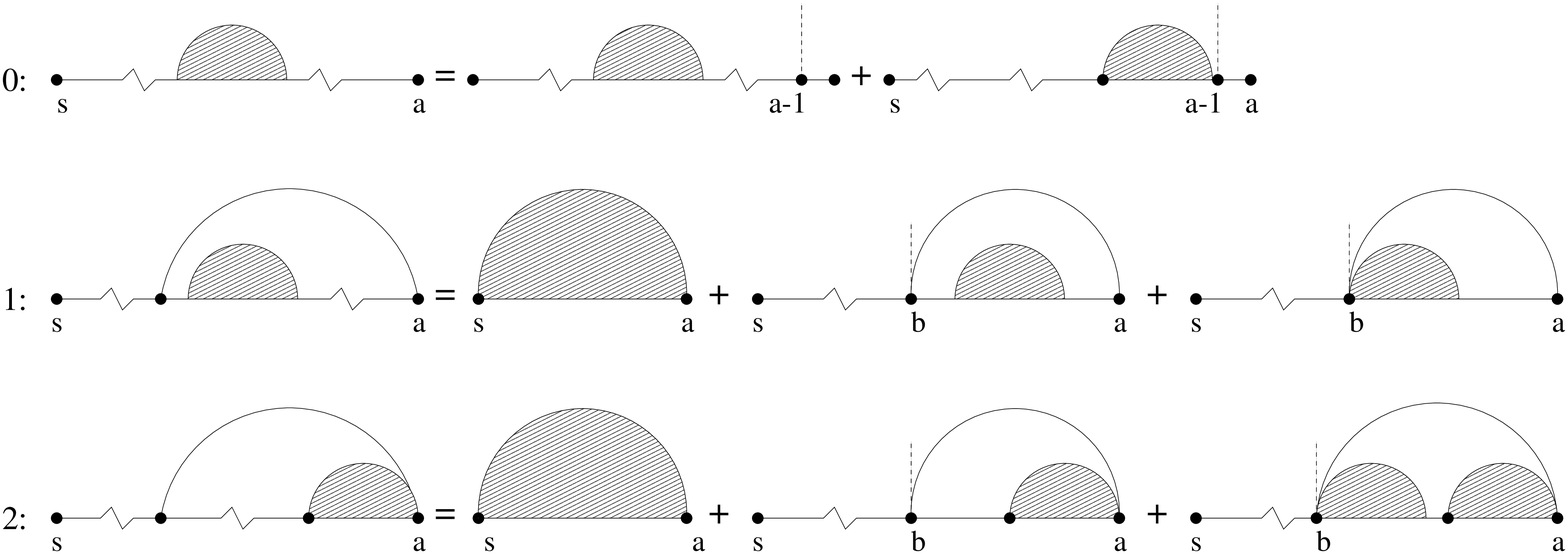}
\caption{the recursive relations for matrices ${\bf G}_t[E,n,s,a]$, 
where $t=0,1,2$.}
\label{recurse}
\end{center}
\end{figure}

\begin{eqnarray}
{\bf G}_0[E,n,s,a]&=&{\bf G}_0[E,n-1,s,a-1]+{\bf G}_1[E,n-1,s,a-1],\\
{\bf G}_1[E,n,s,a]&=&
\delta_{n,1}\sum\limits_{t=L,M,I}{\bf G^\ast}_t[E,s,a]\nonumber\\ 
&+&\sum\limits_{0<b<a}\sum\limits_{E_1}{\bf G}_0[E-E_1,n-1,s,b]
\sum\limits_{t=L,M,I}{\bf G^\ast}_t[E_1,b,a]\nonumber\\
&+&\sum\limits_{0<b<a}\sum\limits_{E_1}{\bf G}_1[E-E_1,n-1,s,b]
\sum\limits_{t=M,I}{\bf G^\ast}_t[E_1,b,a],\\
{\bf G}_2[E,n,s,a]&=&
\delta_{n,1}\sum\limits_{t=R,LR}{\bf G^\ast}_t[E,s,a]\nonumber\\
&+&\sum\limits_{0<b<a}\sum\limits_{E1}{\bf G}_0[E-E_1,n-1,s,b]
\sum\limits_{t=R,LR}{\bf G^\ast}_t[E_1,b,a]\nonumber\\
&+&\sum\limits_{0<b<a}\sum\limits_{E_1}{\bf G}_1[E-E_1,n-1,s,b]
{\bf G^\ast}_{R}[E_1,b,a]. 
\end{eqnarray}
Correspondingly, $g_N(E)$ is extended to ${\cal G}_N[E;f]$ as 
\begin{eqnarray}
{\cal G}_N(E;f)=\sum\limits_{n=1}^{N} \sum\limits_m C_x(n;m) e^{\beta fm}
\sum\limits_t{\bf U}\cdot{\bf G}_t[E,n,s,l]\cdot{\bf U^{t}},
\label{secstrusumnew}
\end{eqnarray}
where $C_x(n;m)$ is the number of conformations of $n$-step OSAWs whose final 
$x$ coordinates are $m$. Eq.~\ref{secstrusumnew} clearly separates 
contribution of the unrelated CGs from the single-stranded parts. 

Because hairpin is one of four elements in secondary structure, ECA 
should be suitable to it. We 
also note that exact enumeration still is available when the lengths of two 
tails are smaller than a given value ($27$ in this paper), a mixture of two 
methods is applied in hairpin case. We believe that it is reasonable, since as 
CG region $(s_o,l_o)$ is so large, or tails are so short that tails 
conformations almost are ``frozen" by excluded-volume interactions, 
enumeration method may be accurate. While the lengths of tails are relatively 
longer than the CG size, or the excluded interactions between CG and tails is 
weaker, the ECA may be available. 

\subsection{Test of ECA without force}\label{testf0}
\begin{figure}[htpb]
\begin{tabular}{cc}
\includegraphics[width=0.4\columnwidth]{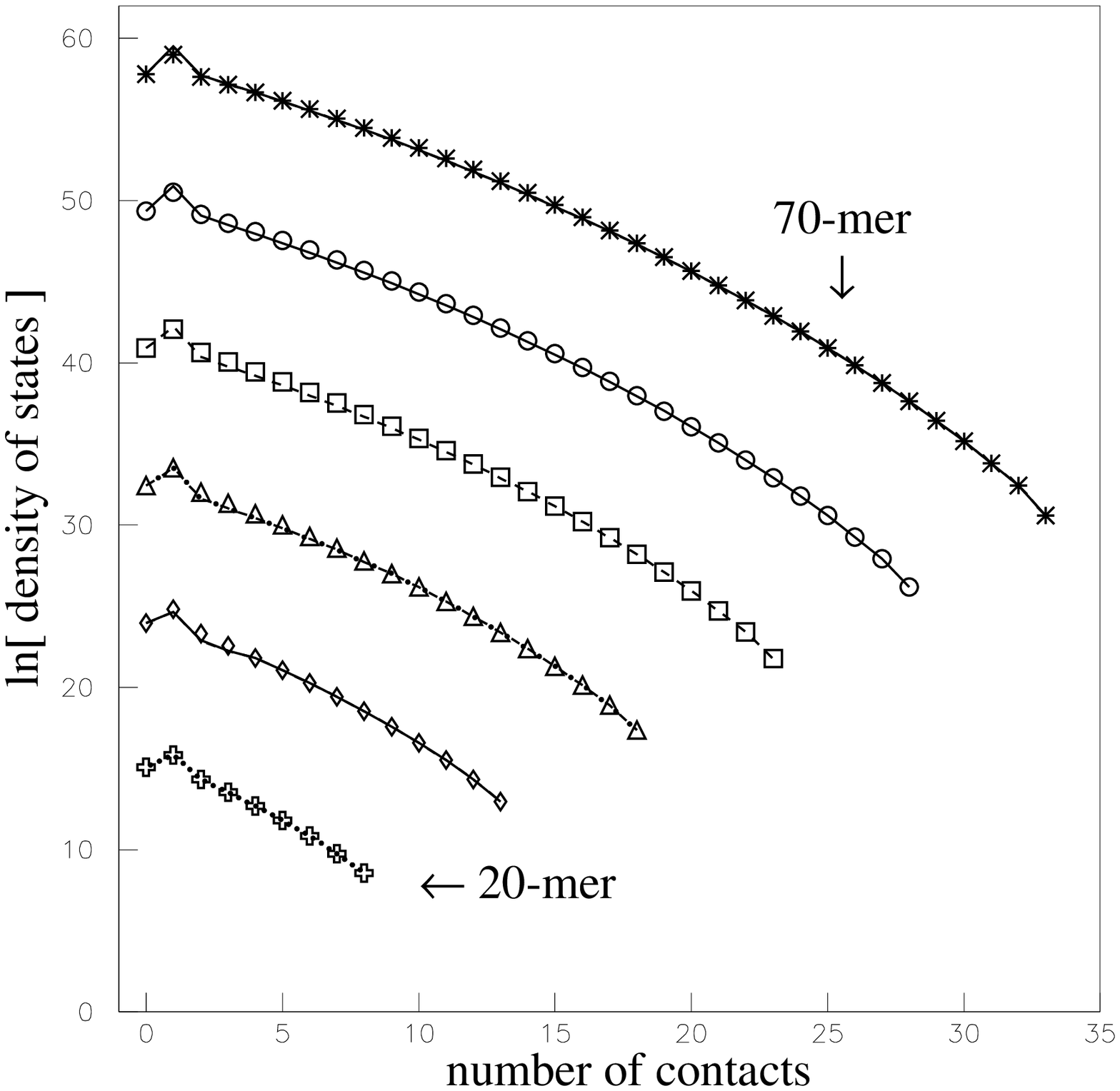}&
\includegraphics[width=0.4\columnwidth]{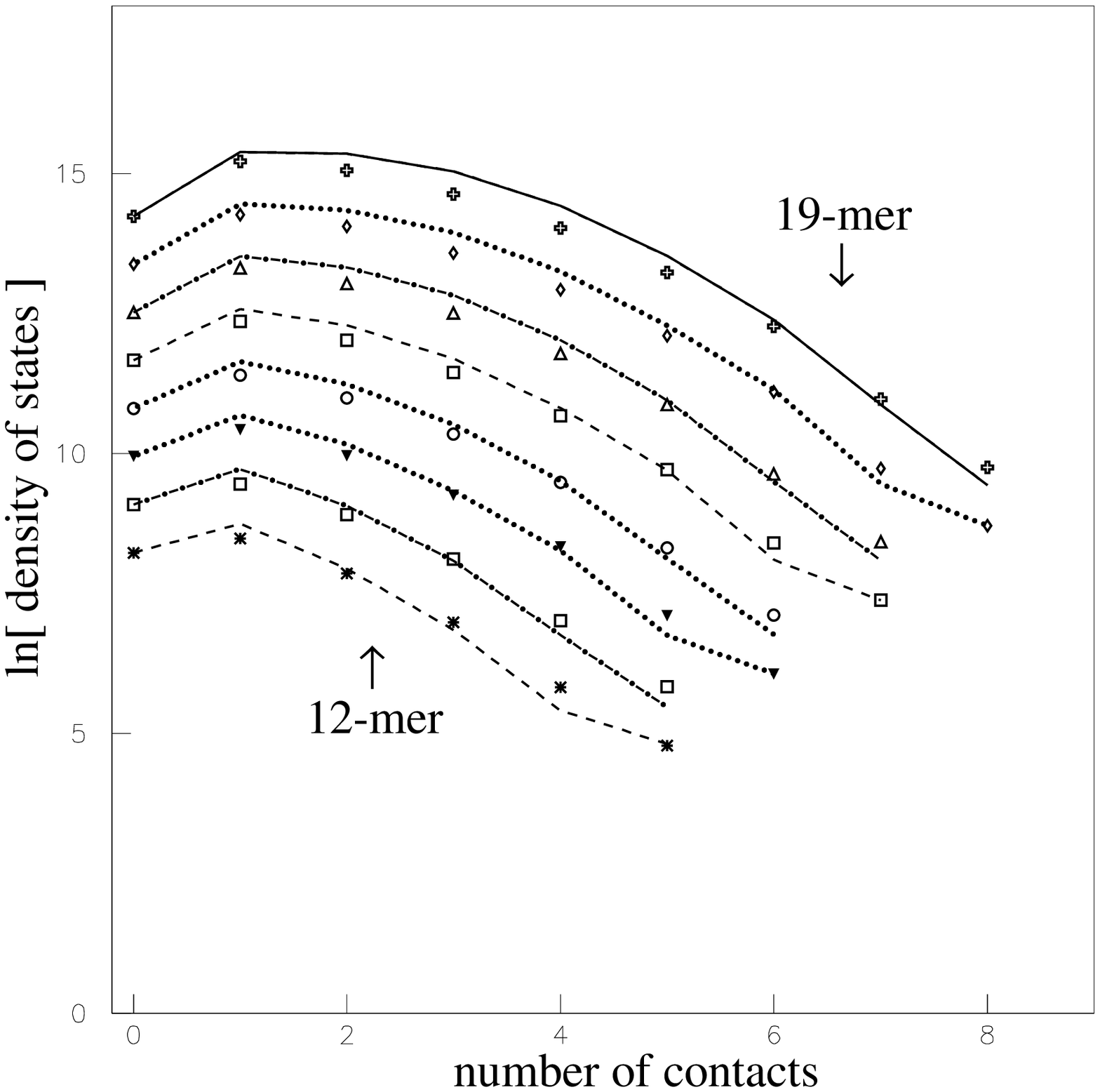}\\
(a)&(b)
\end{tabular}
\caption{(a) Comparing density of states for hairpin conformations 
calculated by enumeration\protect\cite{chen98} method and its mixture with 
ECA, where the symbols come from enumeration method, and the different 
lines represent results from the latter. The energy function used is 
$E=-\varepsilon\times$(number of contacts) ($\varepsilon>0$). The chain monomers 
increase from 20-mer to 70-mer per 10-mer. Two methods give almost the same results. 
(b) Comparing density of states for secondary structure conformations 
calculated by Ref.~\protect\cite{chen98} and ECA, where symbols are from the former, 
and the different lines represent results from ECA 
method. The monomers change from 12-mer to 19-mer. We find that 
our approach gives good approximation.} 
\label{testhairpinsecsnd}
\end{figure}

We compute the density of states for hairpin and secondary structure 
conformations respectively. Comparing with Chen and Dill results, we find that 
our approximation gives considerable consistent results; see 
Figs.~\ref{testhairpinsecsnd}. The results can be tested against exact 
enumeration of shorter chains, which have been showed in Figs. 9 and 14 in 
Ref. \cite{chen98}. 

\section{Tests of the model: EFCs, population distributions 
and re-entering phenomena}
\label{disscussion} 
We have described above how the statistical mechanical model of 
double-stranded chain molecules is extended to force stretching problem. In 
this section, we make use of the model to explore EFCs, 
contact and population distribution, and 
re-entering phenomena. To account for specific monomer sequences, 
sequences are divided into two classes: homogeneous and RNA-like chain. Each 
contact of the homogeneous chain contributes a sticking energy $-\varepsilon$. 
While RNA-like chain has a specific sequence of four types of monomers: A, U, 
C and G, resembling the 4 types of bases of an RNA; only A-U pair or C-G pair 
contributes an sticking energy $-\varepsilon$. In following, we take the lattice 
spacing $b$. 

\subsection{Extension-force curves}
\begin{figure}[htpb]
\begin{tabular}{cc}
\includegraphics[width=0.4\columnwidth]{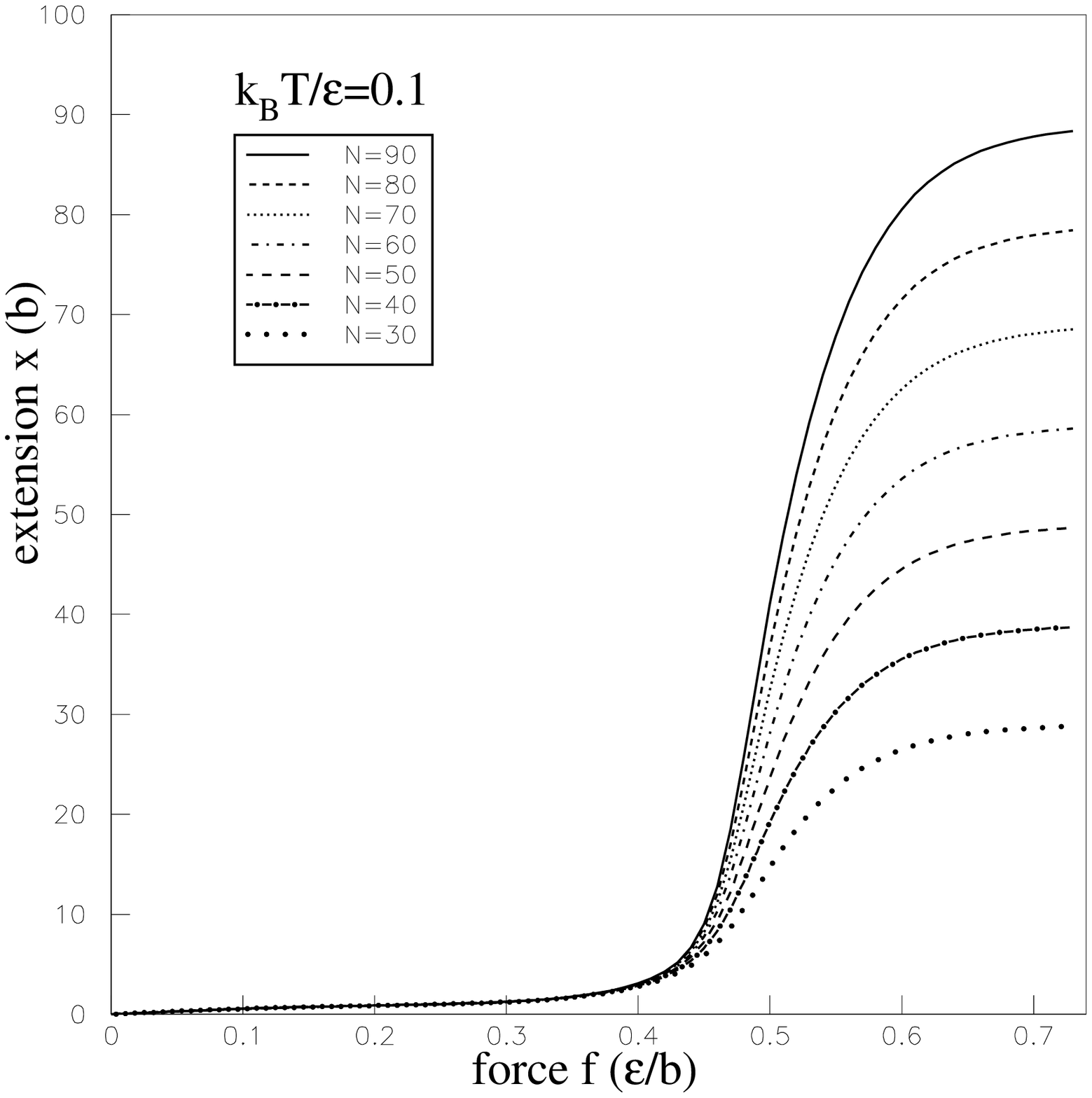}&
\includegraphics[width=0.4\columnwidth]{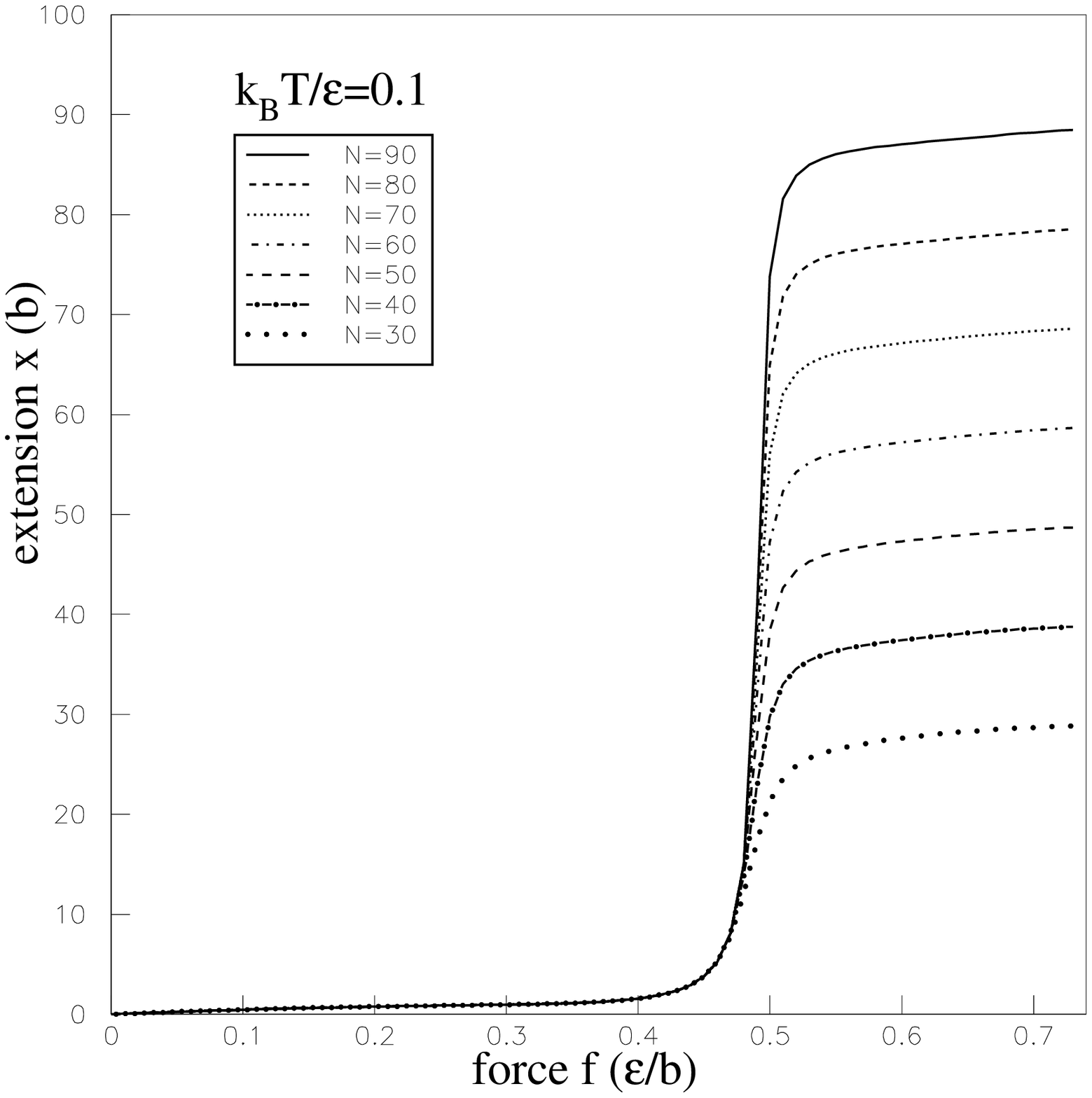}\\
(a)&(b)\\
\includegraphics[width=0.4\columnwidth]{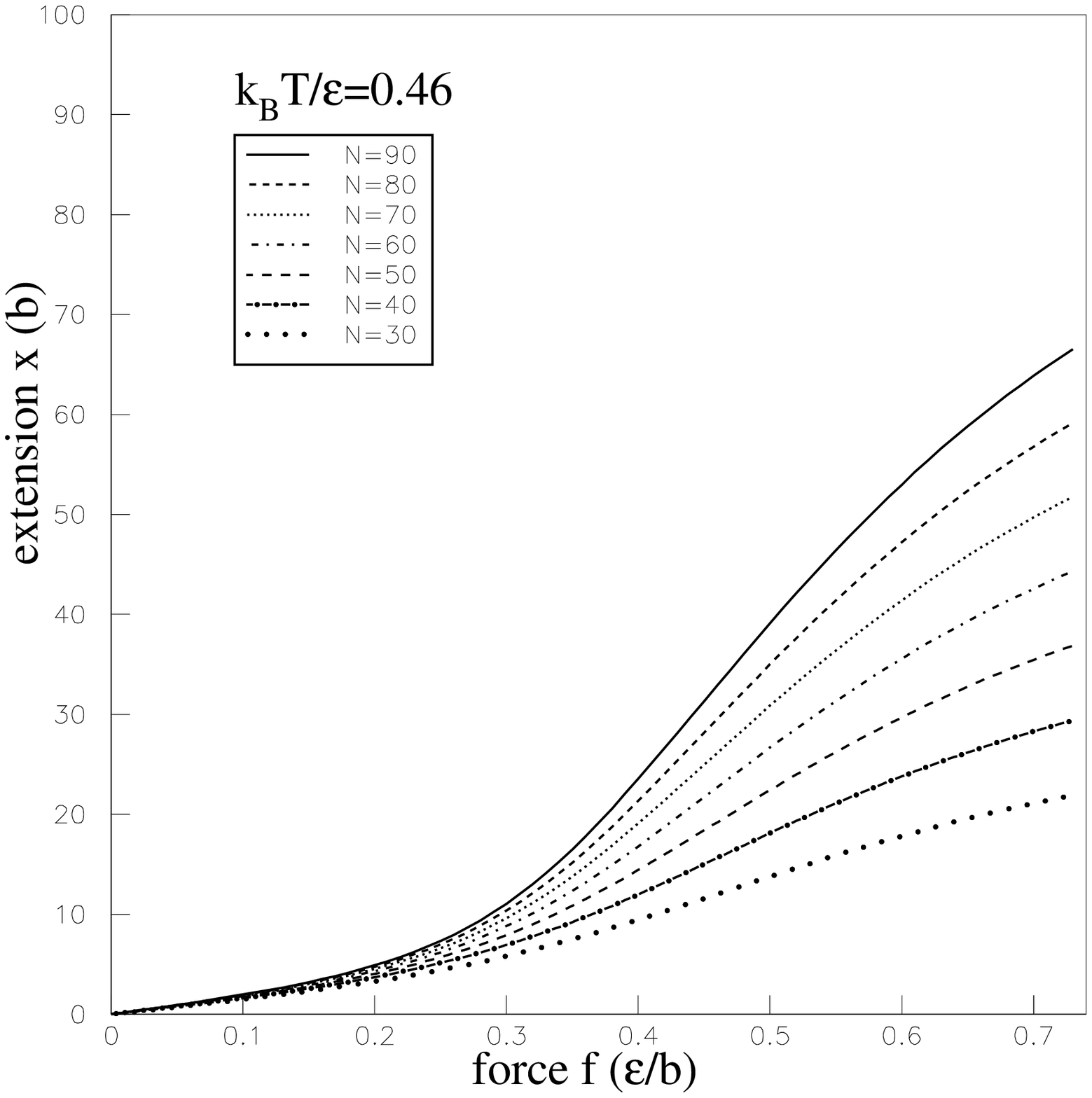}&
\includegraphics[width=0.4\columnwidth]{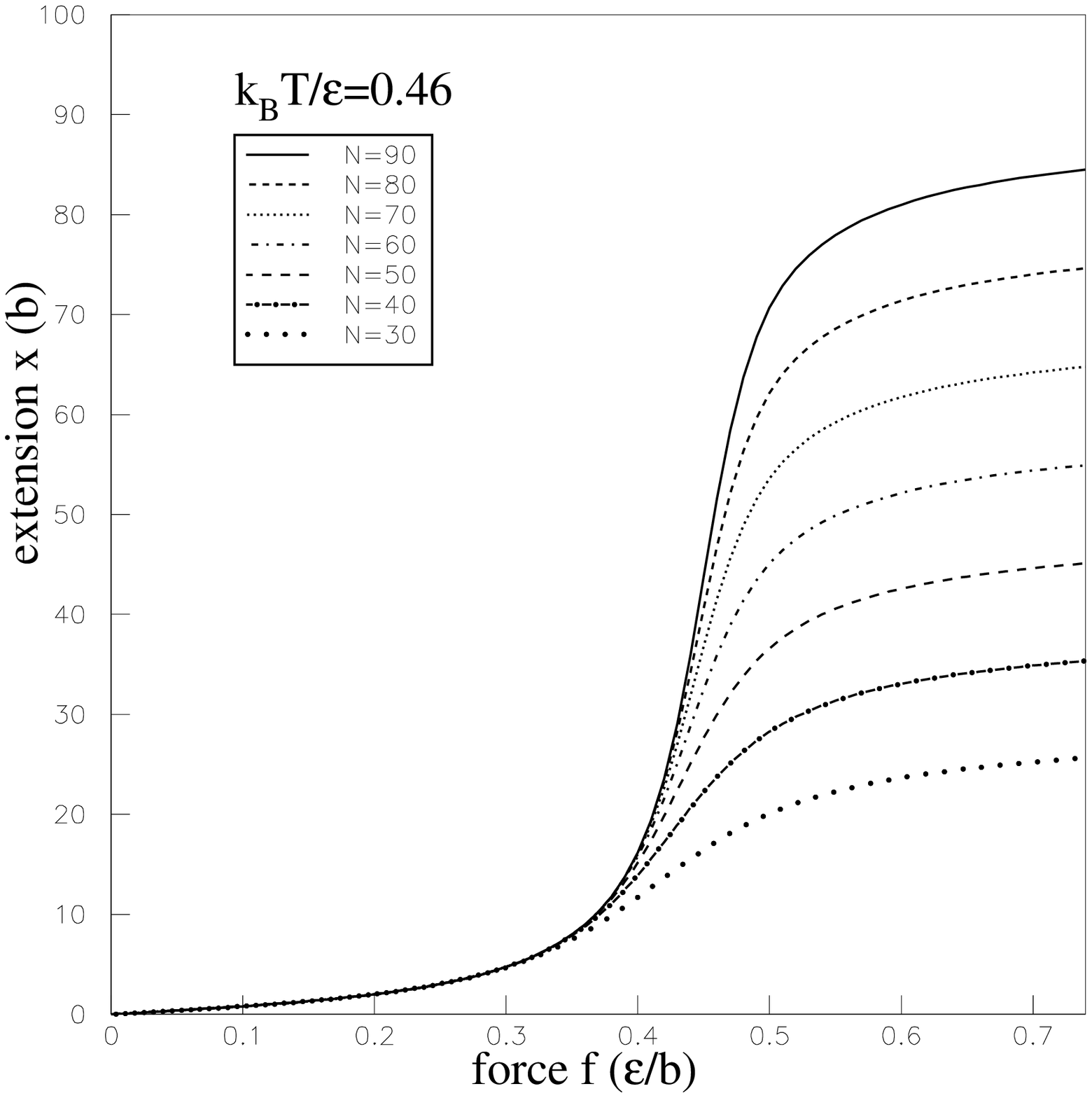}\\
(c)&(d)
\end{tabular}
\caption{The EFCs of homogeneous chains restricted to 
secondary structure conformations: (a) and (c), and hairpin conformations: 
(b) and (d). EFCs of hairpin case have first-order-like behavior being 
similar with unzipping of dsDNA. In contrast, EFCs of secondary structure 
change slowly and smoothly.}
\label{homopolymerefcs01}
\end{figure}

We first investigate how temperature $T$ and monomer number $(N+1)$ affect 
EFCs of homogeneous chains restricted to hairpin and secondary structure 
conformations, respectively; see Fig.~\ref{homopolymerefcs01}. We have 
following results: at given temperature, EFCs of any class 
conformations have similar shapes for any $N$-value; shapes of the EFCs become 
smooth when temperature increases. In particular, extensions in hairpins 
increase sharply to full length during a narrow force range when force 
reaches critical value such as $f=0.48\varepsilon/b$ at 
$T=0.1\varepsilon/k_B$. It is a typical all-or-none behavior. Whereas at the same 
temperature, extensions in secondary 
structures increase at force $0.48\varepsilon/b$, but they reach the full 
length slowly and smoothly until force is about $0.75\varepsilon/b$.  
In addition, the EFCs of hairpins are more stable with temperature than 
that of secondary structures. Generally, 
the dramatical jumps in hairpin EFCs show that contacts in this conformations 
have higher cooperativity. In contrast, 
for secondary structure conformations, there is 
a more larger number of suboptimal states, by which dramatical EED changes 
may be absorbed; lower cooperativity is reflected from smoothness of EFCs. 
The same reason may also explain the different thermal stability of EFCs 
with temperature observed in different structure restriction. 

\begin{figure}[htpb]
\begin{tabular}{cc}
\includegraphics[width=0.4\columnwidth]{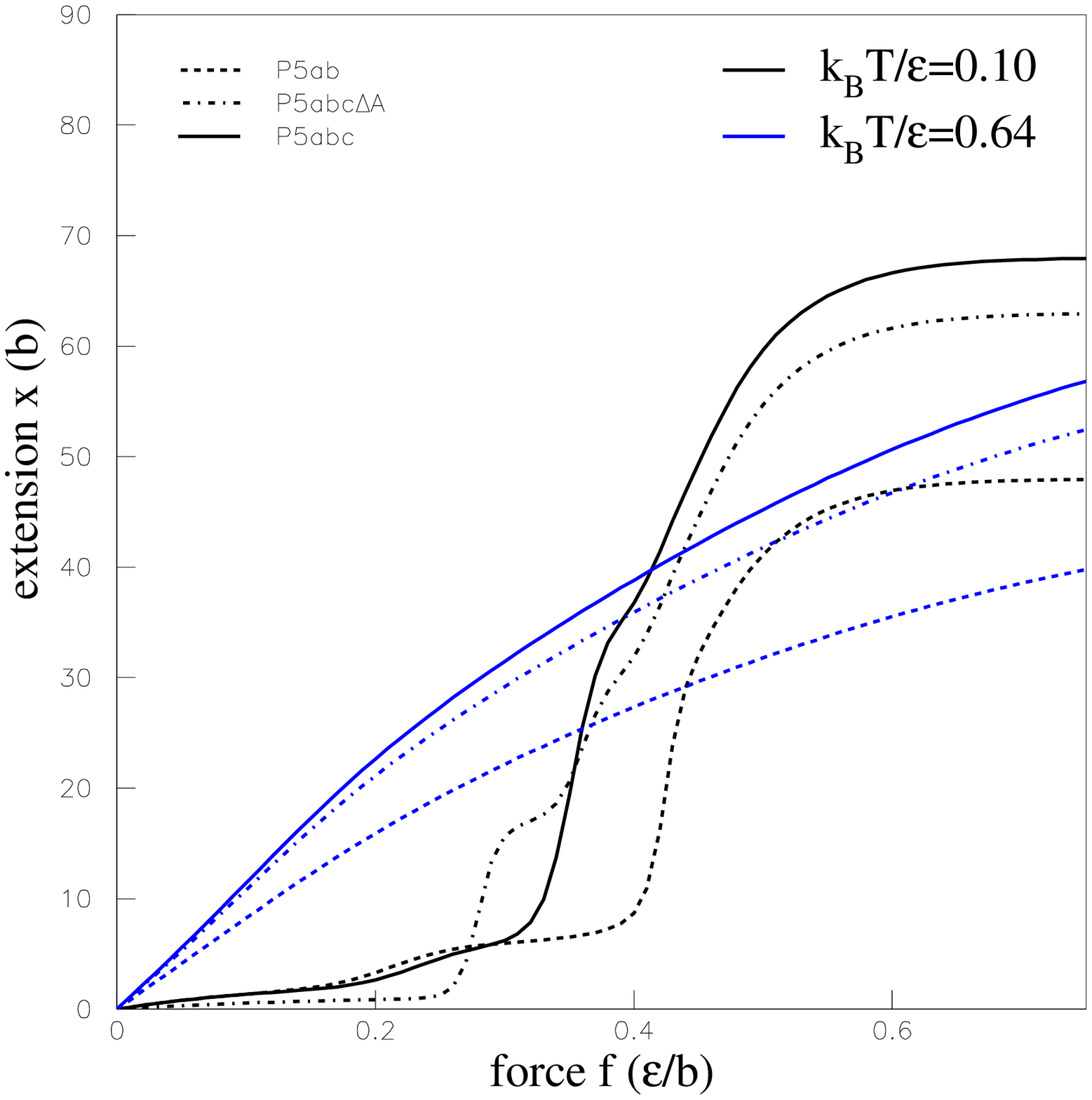}&
\includegraphics[width=0.4\columnwidth]{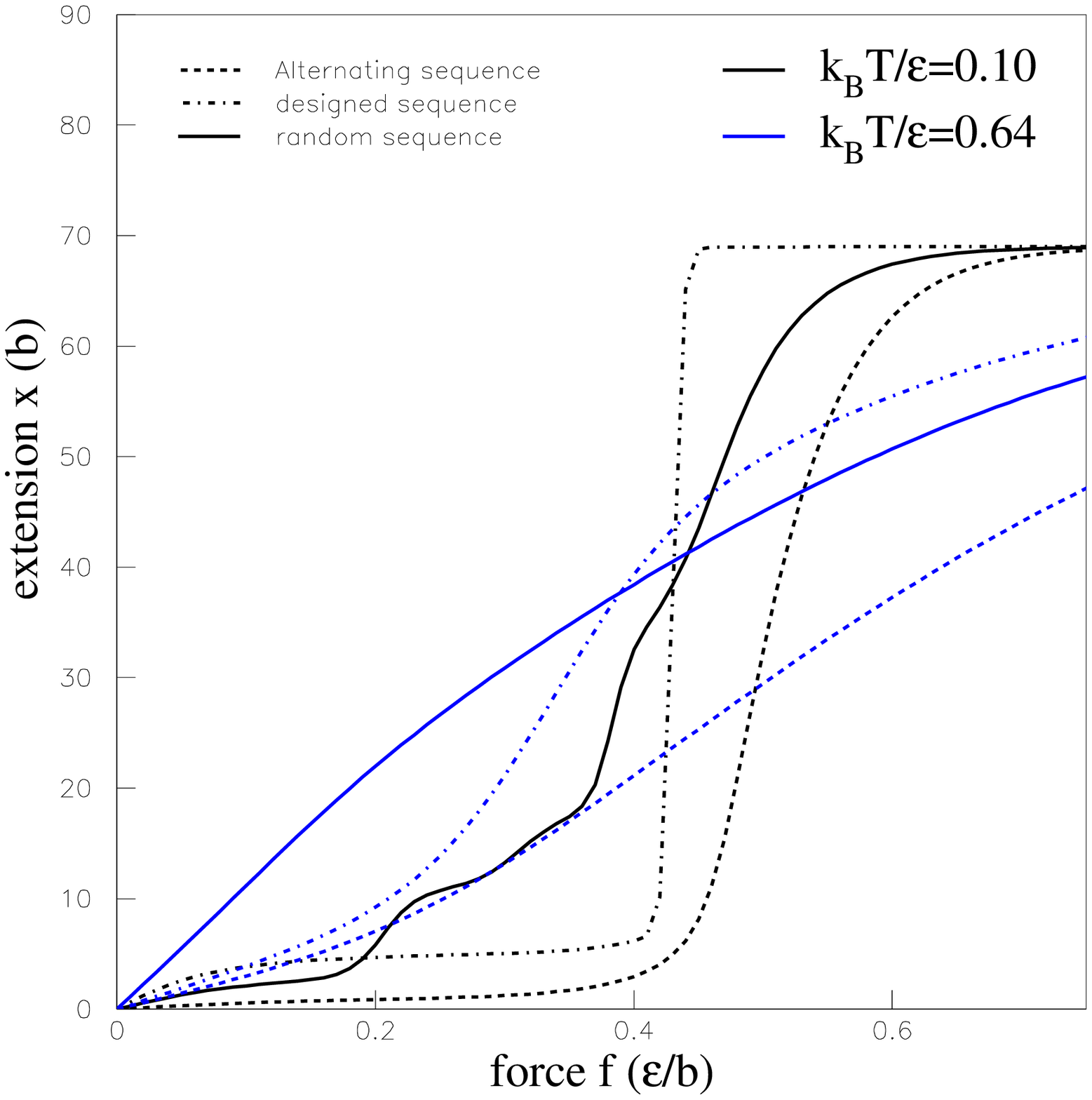}\\
(a)&(b)
\end{tabular}
\caption{The EFCs of heterogeneous secondary structure chains: (a) the 
sequences are 49-mer P5ab, 64-mer P5abc$\Delta$A and 69-mer P5abc. 
(b) 70-mer sequences are random, alternating and designed. Two colors represent 
EFCs at different temperature. }
\label{p5abefcs}
\end{figure}

To illustrate effects of specific monomer sequence to EFCs, we compute EFCs 
of biomolecule sequences P5ab, P5abc$\Delta$A, and P5abc; 
see Fig.~\ref{p5abefcs}(a). These sequences come from force unfolding RNA 
experiment studied by Liphardt {\it{et al.}} recently\cite{liphardt}. 
Thought we do not attempt to fit experimental EFCs in this paper, EFCs of 
secondary structure conformations derived  
from our model do reveal similar information with observed in experiment: 
at $T=0.1\varepsilon/k_B$, unfoldings of P5ab and P5abc start at different 
forces; EFC of P5abc has an apparent intermediate, i.e., after a large jump 
without intermediates at $f=0.37\varepsilon/b$ there follows a inflection. 

We also calculate EFCs of random, alternating and designed sequences; see 
Fig.~\ref{p5abefcs}(b). The designed is composed of four bases, A, U, C 
and G arranged by $A\cdots ACCCCU\cdots UC\cdots CAAAAG\cdots G$, where the 
dots represent 15-mer A, U, C and G consecutively. Comparing the EFCs of 
the artificial sequences with curves got before, we find the EFC of the 
random sequence is very similar with the EFCs of biomolecular sequences in 
Fig.~\ref{p5abefcs}(a); 
the EFC of the alternating sequence is almost the same with the results of 
homogeneous chain of secondary structure; while the EFC of the designed 
sequence like the EFCs of homogeneous hairpins. From designed sequence 
arrangement, we think that the chain prefers to form two hairpins 
simultaneously. So if the force can unzip one of two hairpins, it is also 
large enough to unzip another at the same time, which can be reflected from the 
large jump of the extension from small to full length abruptly. 

\subsection{Monomer-monomer contact distribution and population distribution}
\begin{figure}[htpb]
\begin{tabular}{cc}
\includegraphics[width=0.4\columnwidth]{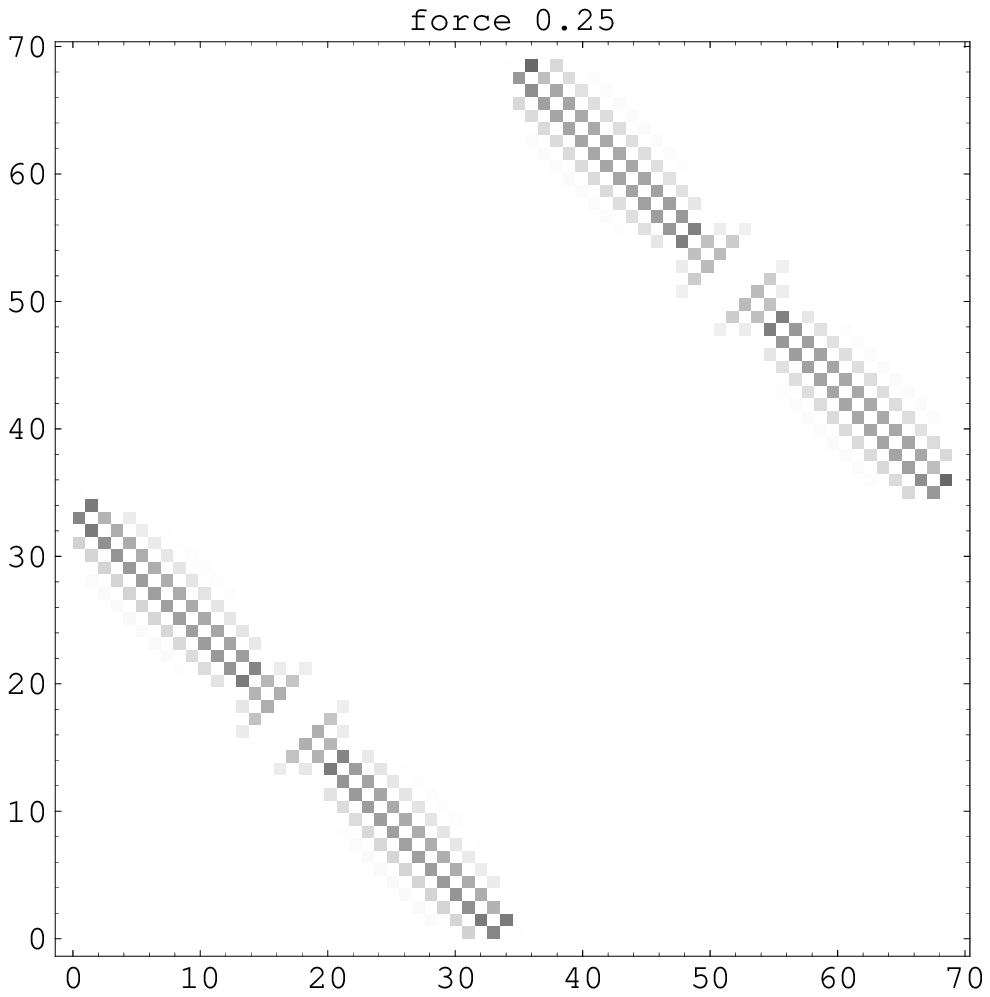}&
\includegraphics[width=0.4\columnwidth]{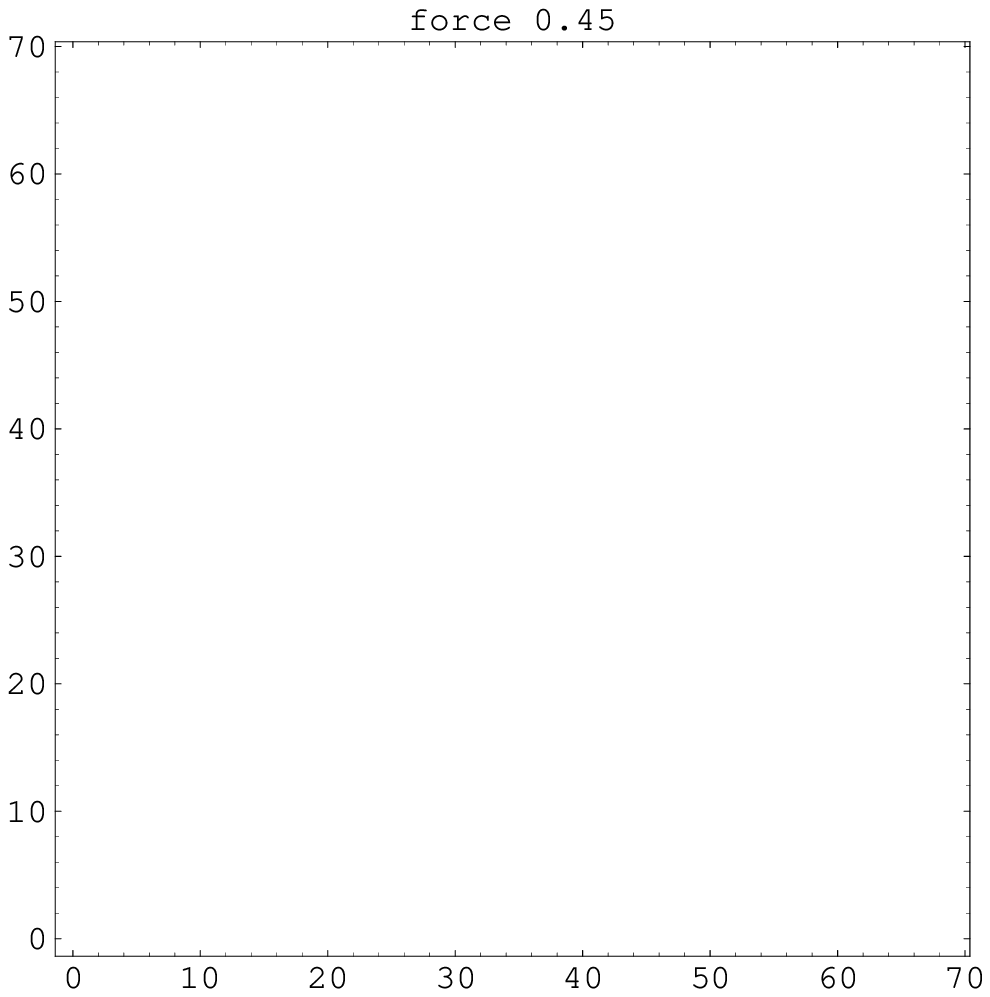}\\
(a)&(b)
\end{tabular}
\caption{The density diagrams for contact probability $p(i,j;T,f)$ of the 
designed sequence restricted to secondary structure conformations, where forces 
are $0.25$ and $0.45\varepsilon/b$ at temperature $0.1\varepsilon/k_B$.
}
\label{density1}
\end{figure}

Our model not only can predict EFCs of different sequences, but also it 
relates the EFCs with structures. We calculate contact 
distributions $p(i,j;T,f)$ for every possible contact pair 
$(i,j)$\cite{chen98}. The equilibrium $p(i,j;T,f)$ is determined by 
the ration of the conditional partition function ${\cal Q}_N(i,j;T,f)$ for 
all the conformations that contain contact $(i,j)$ and the full partition 
function ${\cal Q}_N(T;f)$ defined in Eq.~\ref{engsumforc}, 
$p(i,j;T,f)={\cal Q}_N(i,j;T,f)/{\cal Q}_N(T;f)$. From the contact 
distribution, the structure of molecule is derived at given force and 
temperature. As an illustration, density diagrams of the distributions of 
the designed sequence restricted to secondary structure are shown in 
Fig.~\ref{density1}. It is apparent to see that two hairpins almost present or 
disappear simultaneously according to the force values, smaller or larger 
than $0.42\varepsilon/b$.

\begin{figure}[htpb]
\begin{tabular}{cc}
\includegraphics[width=0.4\columnwidth]{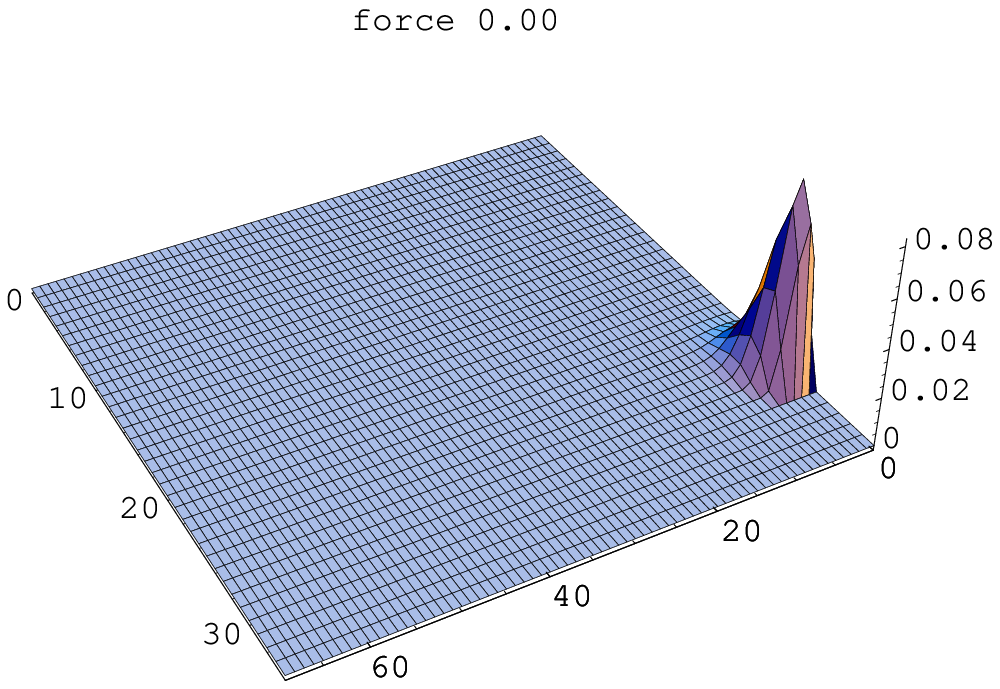}&
\includegraphics[width=0.4\columnwidth]{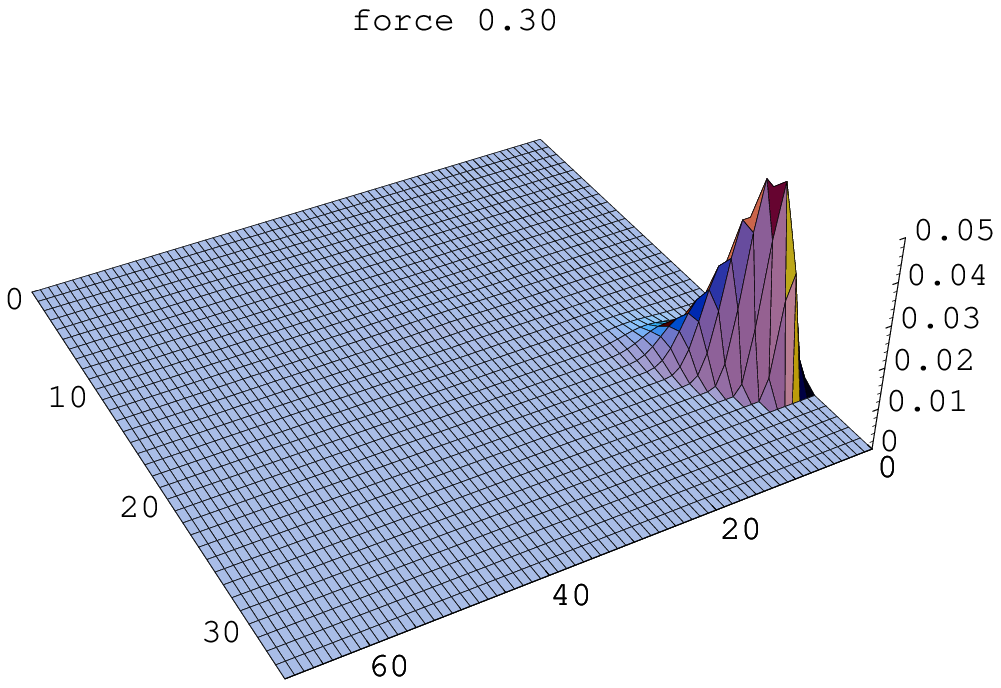}\\
(a)&(b)\\
\includegraphics[width=0.4\columnwidth]{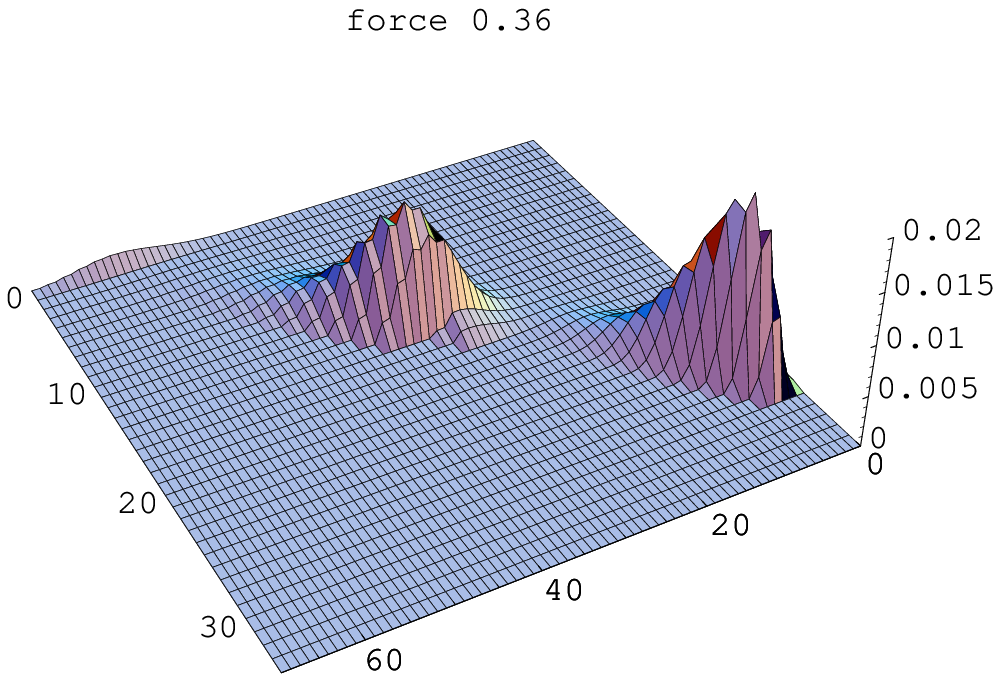}&
\includegraphics[width=0.4\columnwidth]{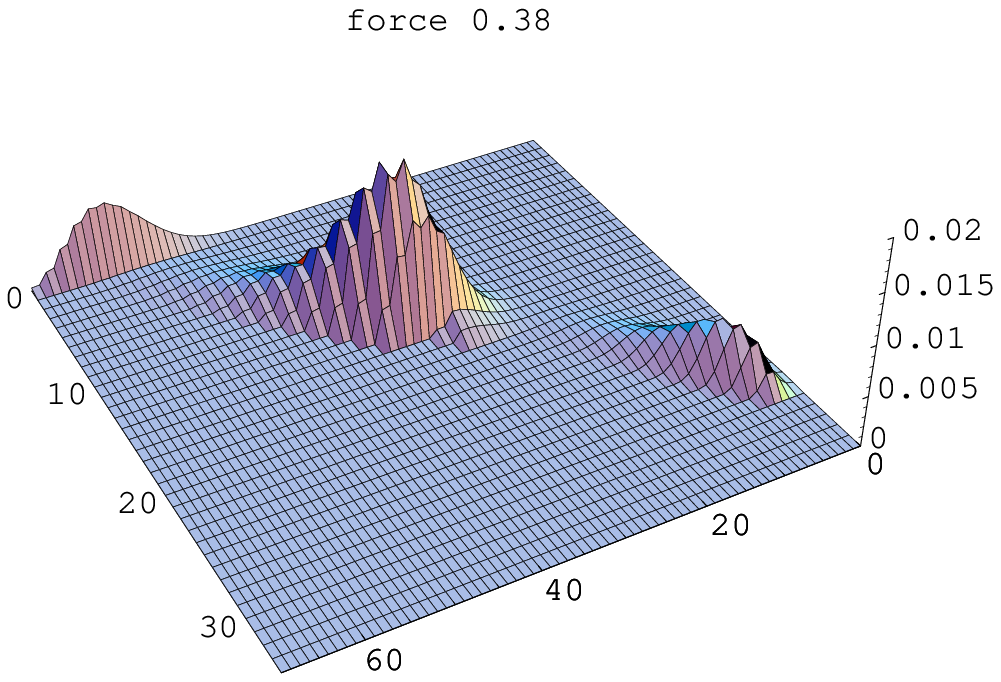}\\
(c)&(d)\\
\includegraphics[width=0.4\columnwidth]{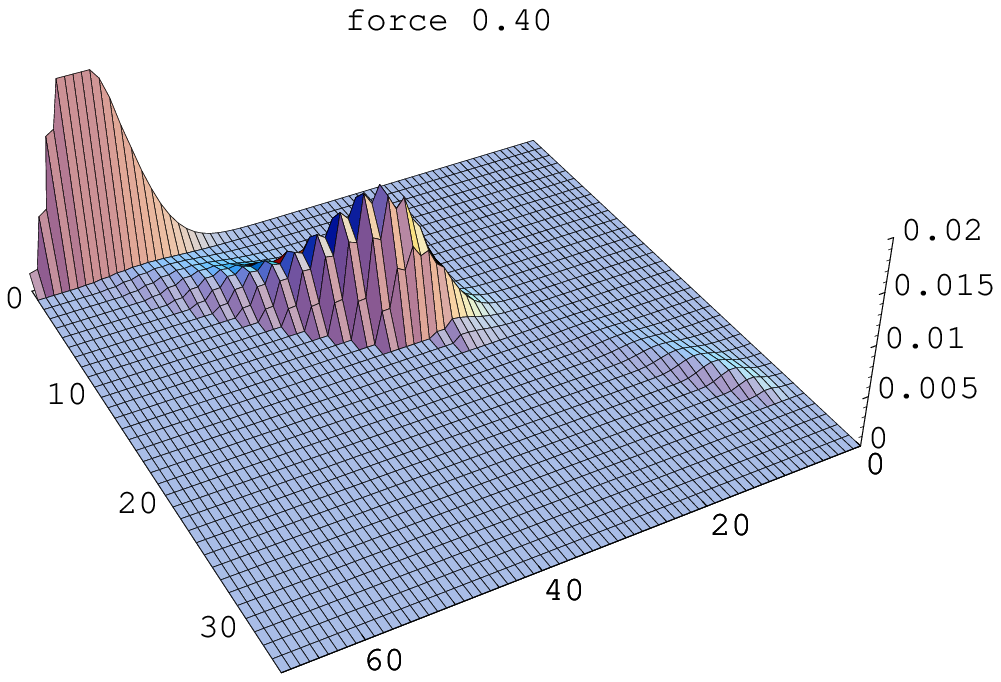}&
\includegraphics[width=0.4\columnwidth]{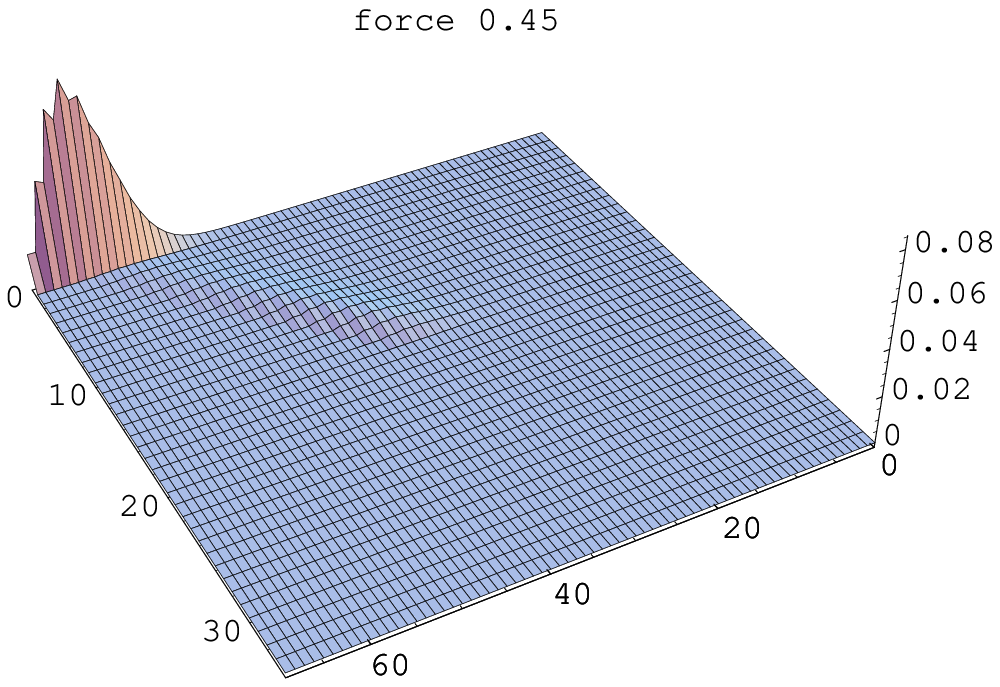}\\
(e)&(f)
\end{tabular}
\caption{The population landscapes $p(\Delta,q;T,f)$ of the designed sequence 
restricted to secondary structure conformation, where forces in unit 
$\varepsilon/b$, the temperature $0.37\varepsilon/k_B$, and the coordinate from 
0 to 70 (x-axis) is the EED $\Delta$, while another coordinate (y-axis) is the 
contact $q$.  }
\label{pathways}
\end{figure}

Although contact distribution provides the information of chain structure, it 
is insufficient to see globally how force unzips complex chain molecules, or what 
role cooperativity plays. In thermal melting RNA, energy 
landscape $F(n,nn)$ was used to provided insights into cooperativity for 
the melting transition, where $n$ and $nn$ are the number of native contacts 
and non-native contacts respectively\cite{chen00}. For our purpose, the 
EED $\Delta$ and total contacts $q$ of conformations are taken as ``order 
parameters".  Instead using free energy surface, we define population 
distribution    
\begin{eqnarray}
p(\Delta,q;T,f)=\frac{1}{{\cal Q}_N(T;f)}{\cal Q}_N(\Delta,q;T,f), 
\label{defpopulate}
\end{eqnarray}
where ${\cal Q}_N(\Delta,q;T,f)$ is conditional partition function 
for all conformations whose EEDs and the number of contacts are $\Delta$ and 
$q$. The probability $p(\Delta,q;T,f)$ carry the same information as the 
free energy. As an application, the population landscapes of the designed sequence 
with six typical forces at the same temperature $0.37\varepsilon/k_B$ are shown 
successively in Fig.~\ref{pathways}. At a force smaller than 
$0.33\varepsilon/b$, virtually all conformations populate the ``native" 
structure, in which two hairpins coexisting. Increasing force above 
$0.33\varepsilon/b$, the conformations separate into two classes: one 
class is that conformations are still two hairpins coexisting; the other 
class is that only one hairpin exists in all conformations. Increasing 
force further (Fig.~\ref{pathways}(d)), three classes of conformations present: 
except two mentioned above, the unfolded or full stretched conformations form a 
new class. Finally Fig.~\ref{pathways}(f) shows that all molecules are full 
unfolded as force is larger than $0.43\varepsilon/b$.  
 
According to Eq.~\ref{extenforce}, the derivative of extension $x$ with 
respect to force $f$ can be expressed as $d<x>/df=(<x^2>-<x>^2)/k_BT$. 
The formula is almost the same with the heat capacity $C(T)$, except that  
extension $x$ is replaced by energy, and force is replaced by temperature. 
The analogy of heat capacity and EFCs can give lots of insights. E.g., in 
thermal melting proceeding, because the peaks 
in $C(T)$ curves mark molecular structure transitions, the sudden increases 
in EFCs might be also understood as structure transitions driven by external 
forces, while the counterparts of the melting temperatures at peaks in $C(T)$ 
are the ``melting" forces at large jumps in EFCs. Based on the same reason, 
conformations between two extension jumps could be called in 
``intermediate" states like in thermal melting case, such as predicted in 
Fig.~\ref{p5abefcs} and observed in experiment\cite{liphardt} about P5abc. 
So, just like the important roles of energy played in heat capacity 
concepts, we believe that EEDs may  
play the similar roles in force stretching problem, at least in nucleic acids. 
The most direct proof is that the population distribution 
$p(\Delta,q;T,f)$ of EED and energy in conformations are 
closely-related, as is showed in Fig.~\ref{pathways}. 

\begin{figure}[htpb]
\begin{tabular}{cc}
\includegraphics[width=0.4\columnwidth]{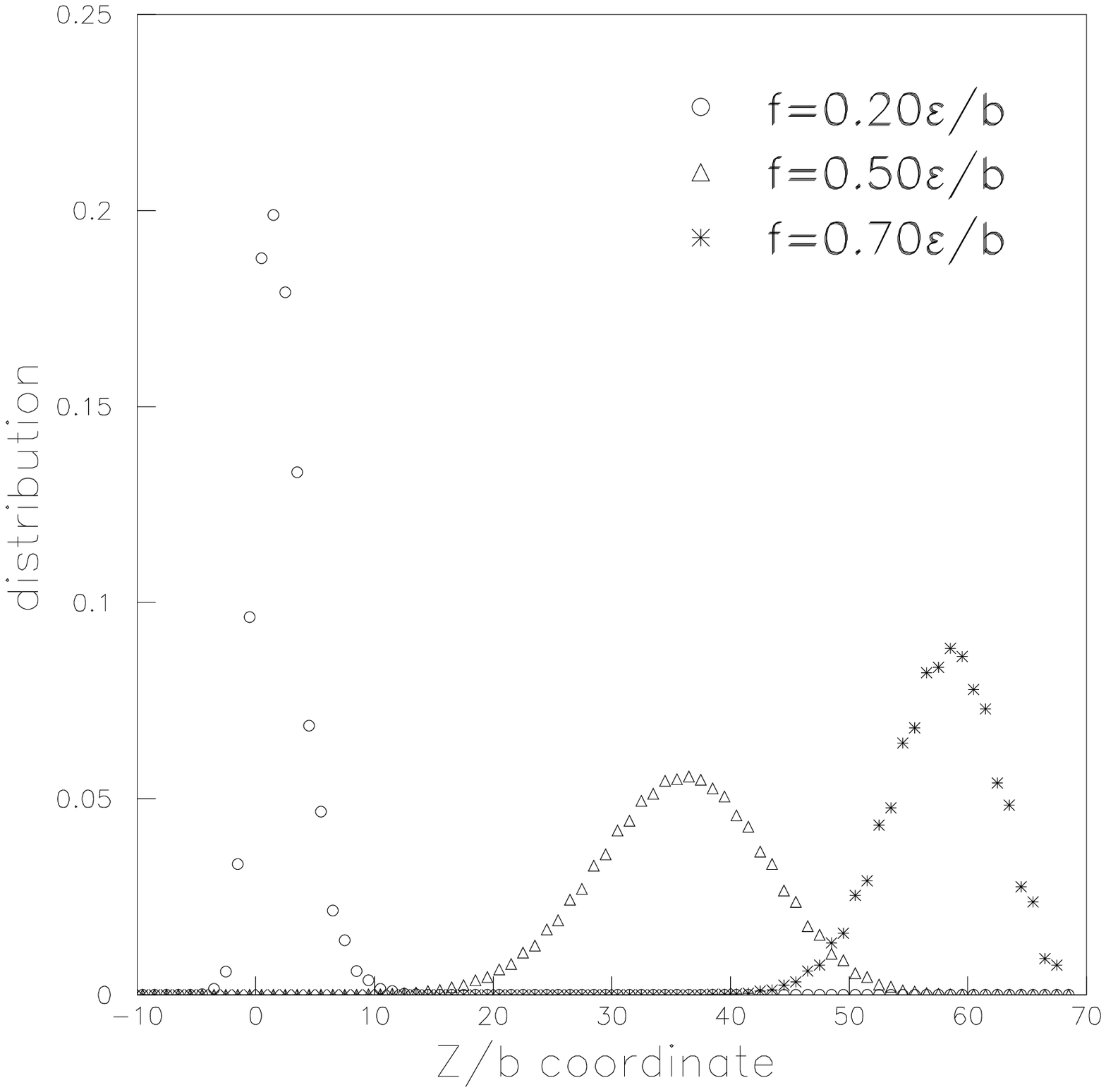}&
\includegraphics[width=0.4\columnwidth]{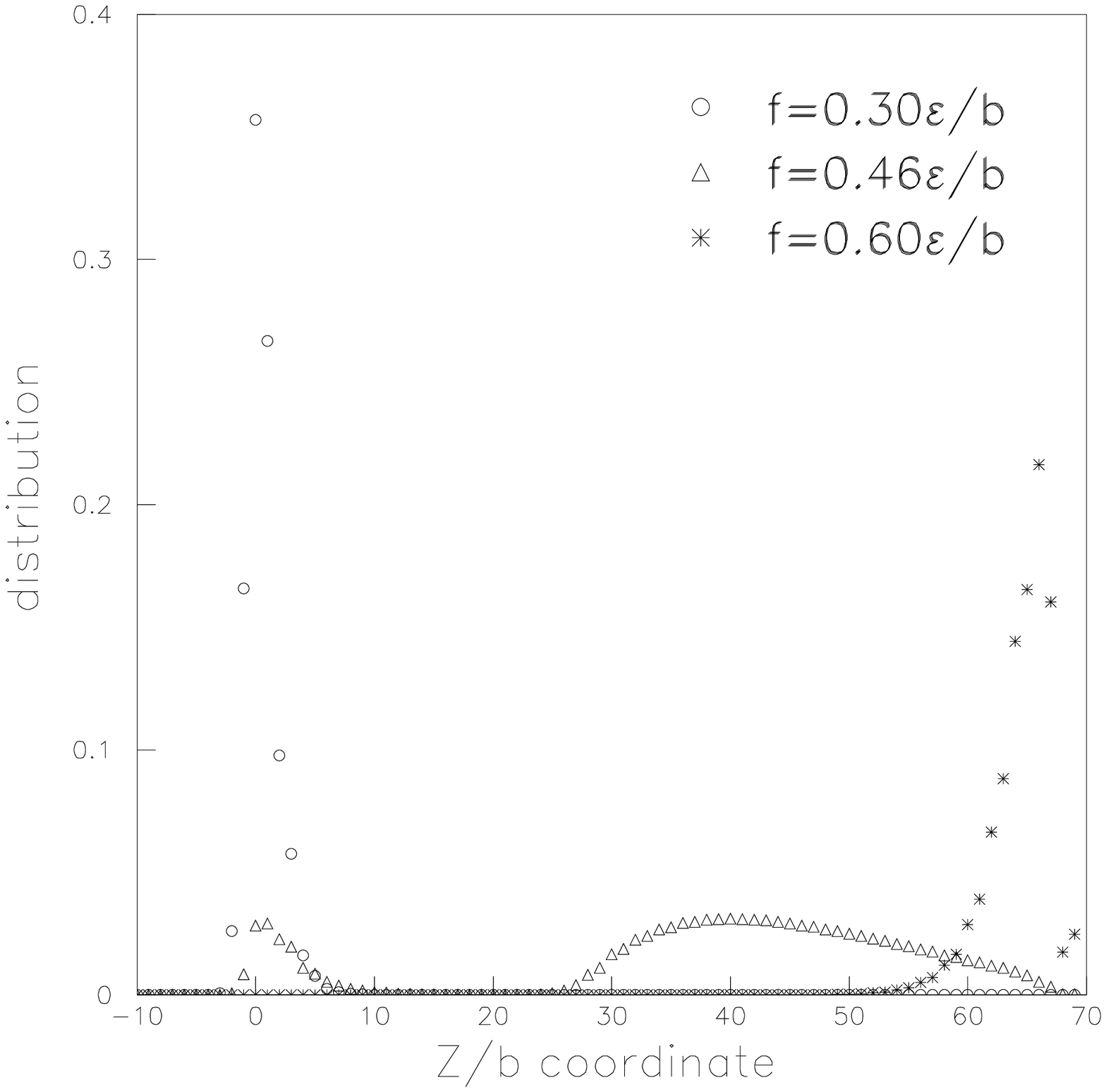}\\
(a)&(b)
\end{tabular}
\caption{The EED distribution $p(\Delta;T,f)$ of the homogeneous chains of 
secondary structure (a) and hairpin conformations (b), where different forces 
are in unit $\varepsilon/b$, and the temperature is $0.25\varepsilon/k_B$. 
At force $0.46\varepsilon/b$, two isolated peaks appearing in the 
distribution represent two-state transition in hairpin conformations. }  
\label{eedd}
\end{figure}

The corresponding of the EED and energy is not needless. To 
show its usefulness, we explore the EED distribution of the chains 
of hairpin and secondary structure conformations. The EED 
distributions $p(\Delta;T,f)$ can be given by a sum over contacts 
$q$ of $p(\Delta,q;T,f)$, 
\begin{eqnarray}
p(\Delta;T,f)=\sum_{q}p(\Delta,q;T,f). 
\end{eqnarray}
Fig.~\ref{eedd} is the distributions of the 70-mer homogeneous double-stranded 
chains: for the secondary structure conformations, only one maximum 
is observed at any force; while in the hairpin conformations, at some forces, 
here near the melting force $0.46\varepsilon/b$ at temperature 
$0.25\varepsilon/k_B$, 
two distinct populations of shorter and longer extensions present, but no 
conformations with other lengths in between. We name two structure 
transitions as ``one-state" and ``two-state", respectively. The transition 
definition is borrowed from thermodynamics of polymers, in 
which the distribution is about molecular microscopic energy\cite{tostesen}. 
The one-state and two-state transitions are resemble a critical 
point and ``first-order transitions" in thermodynamic limit. The 
difference of EED distributions 
of two classes conformations warns that the simple EFCs may cover critical 
physical informations. Like the importance of the width of melting peaks in 
determining thermal transitions type, the third-order derivative of the average 
extension with respect to force is essential to determine the transitions types 
in the force ``melting" double-stranded chain molecules.  

To be different from ordinary polymers, mechanical properties of 
biomolecules are more complex. The main reason is the effect of monomer 
sequence\cite{lubensky,muller225,muller348}. 
Hence, the types of structure 
transitions driven by force is inevitable to be affected by sequence. The 
apparent example is the EED distribution of the designed sequence, which 
can be see explicitly from Fig.~\ref{pathways}: at temperature 
$0.37\varepsilon/k_B$, three isolated peaks in the distribution at 
force $0.38\varepsilon/b$, two peaks at force $0.36$ and 
$0.40\varepsilon/b$. In fact, EFC of the sequence at the temperature 
seems very trivial; the curve is only more smoother than the EFC at 
$0.1\varepsilon/b$.

\subsection{Re-entering phenomena}
\begin{figure}[htpb]
\begin{tabular}{cc}
\includegraphics[width=0.4\columnwidth]{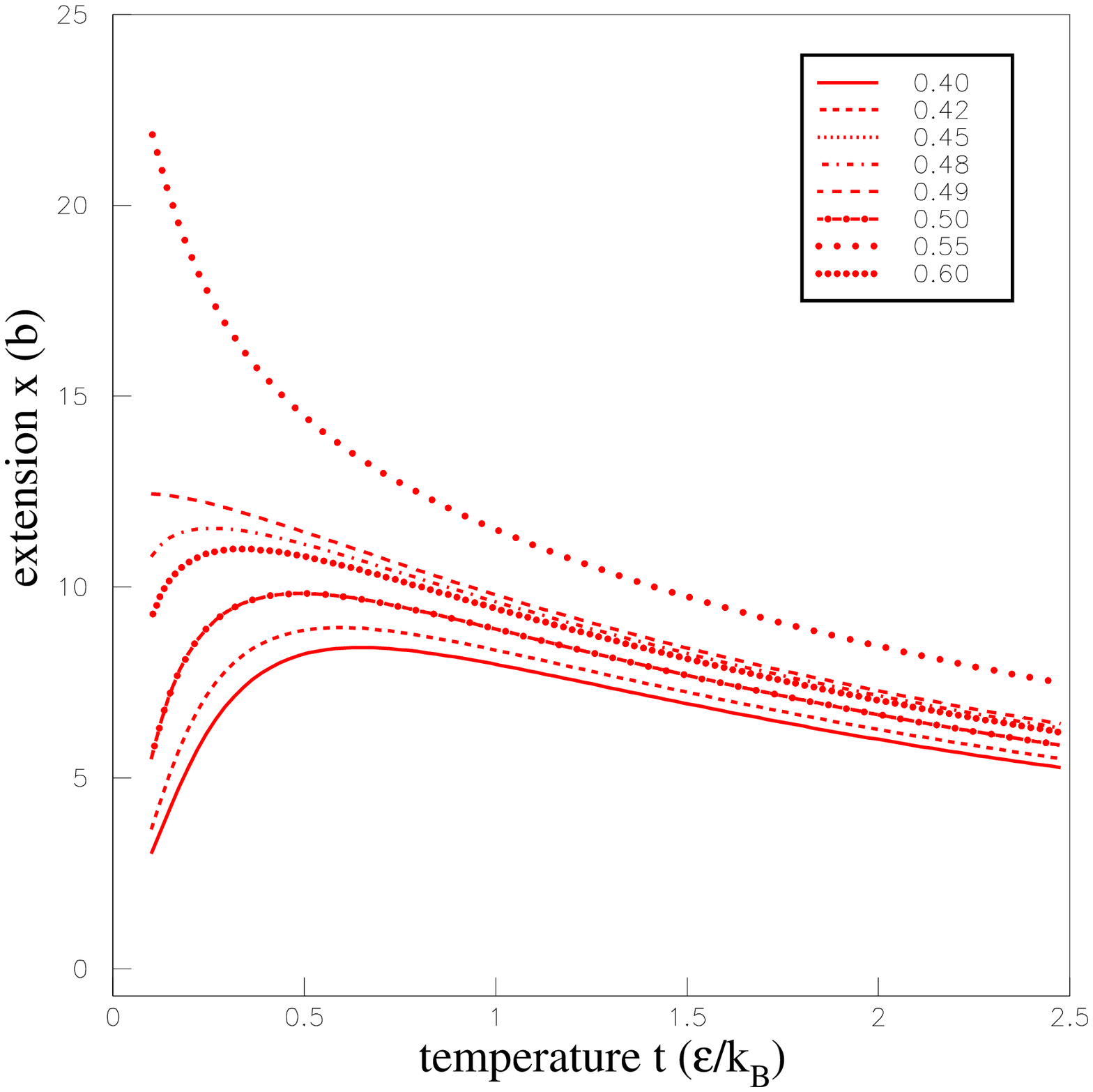}&
\includegraphics[width=0.4\columnwidth]{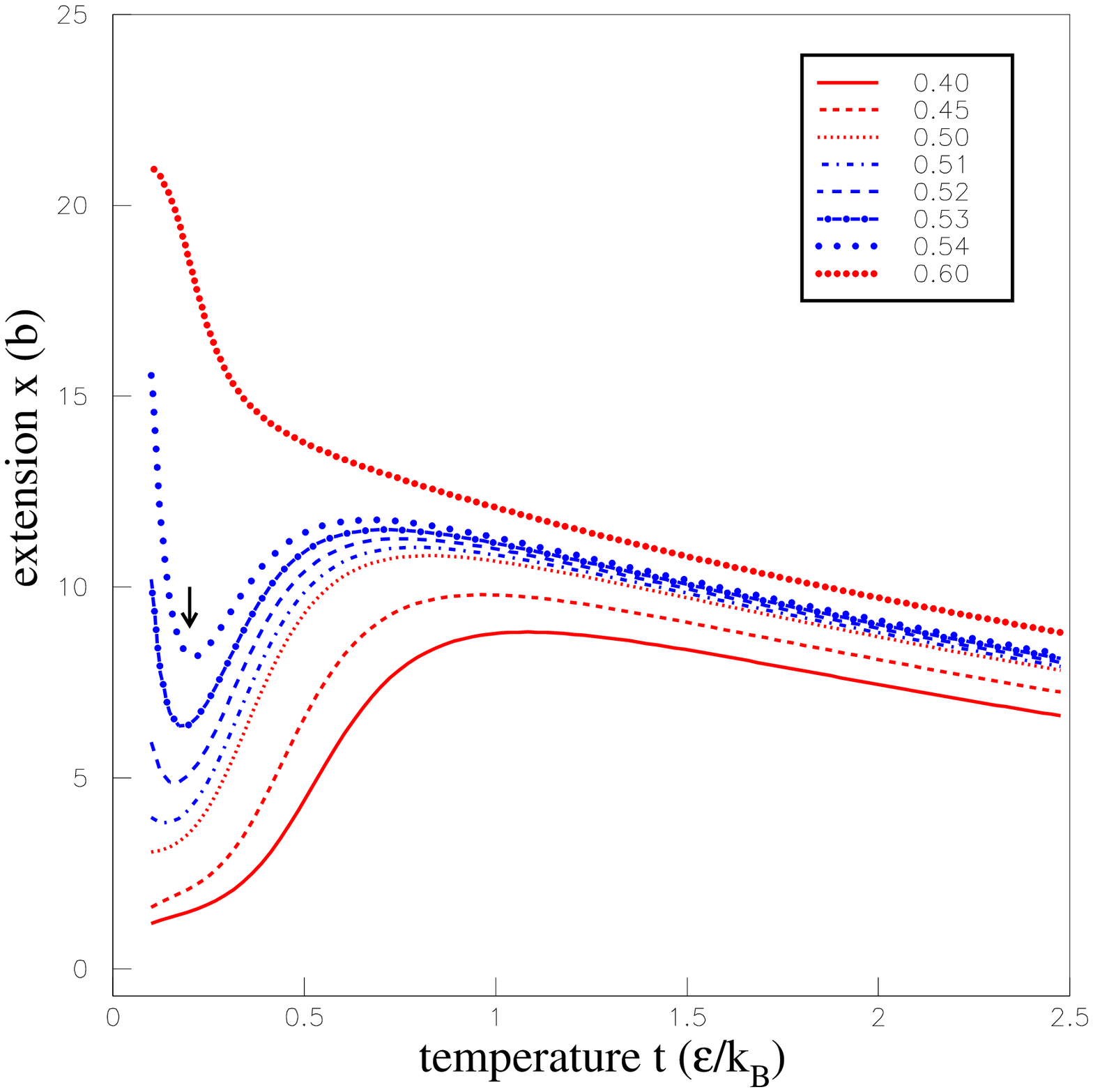}\\
(a)&(b)
\end{tabular}
\caption{The extension-temperature curves for 25-mer chains of 
secondary structure (a) and hairpin conformations (b) at different 
force values, where the sequences are homogeneous. Arrow in (b) points 
out dip minimas, which demonstrates the appearance of re-entering 
transitions in hairpins. }
\label{reenter}
\end{figure}

Finally, we explore the extension dependence on temperatures as external 
force fixed. The studies are related with the phenomena about re-entering 
transitions happened at low temperature in stretching dsDNA case
\cite{marenduzzo,orlandini}. It means that if one fixes external force at 
a value in a finite range and decreases temperature, beginning with 
stretched state, chain molecule will first collapse to a globular state; 
while temperature is lowered further, chain will re-enter stretched state 
again. One can judge the 
existence of the phenomenon from extension-temperature curves directly, 
i.e., there are dips at lower temperature on the curves for some force 
values. Recently M$\ddot{u}$ller pointed out that the re-entering phenomenon 
also present in force stretching RNA molecules\cite{muller225}. It is 
interesting to see whether our model can predict the transitions. We check 
it by numerical computing for homogeneous chains of secondary structure and 
hairpin conformations respectively. These results are shown in 
Fig.~\ref{reenter}, The numerical data demonstrate that the re-entering 
transitions only present in hairpin conformations, but are not been 
found in secondary structure.

\section{Summary}
\label{conclusion}
In this paper, we modify and extend the statistical model of 
double-stranded chain molecules developed by Chen and Dill to force 
stretching case. Unlike several theoretical models proposed previously, 
specific monomers sequence and excluded volume interaction are exactly 
included. We use our model to investigate how EFCs depend on 
chain length, sequence, and structure. In addition, our 
model can relate EFCs with molecular structure directly. And we also 
investigate re-entering phenomenon in hairpin and RNA conformations. 
Because our aim in this paper is just to illustrate, our model is largely 
simplified, e.g., the chain is restricted on 2D lattice, the chain 
stiffness (bending) and elasticity are neglected, and no stacking 
interaction is involved which may be important in force stretching 
theory\cite{zhang}. However, our results still reveal many important 
and interesting physical results. In future work, more realistic 
requirements, such as eliminating of lattice restriction\cite{chen98} 
and adding of stacking energy could be considered.

\begin{acknowledgments}
We are indebted Dr. Shi-jie Chen, Tao Xu, and Prof. H.-W. Peng for many helpful 
discussions in the work process.
\end{acknowledgments}

\appendix
\section{Calculating end-to-end distance distribution of hairpin conformations} 
\label{endtoendhairpin}

\begin{figure}[htpb]
\begin{center}
\includegraphics[width=1.\columnwidth]{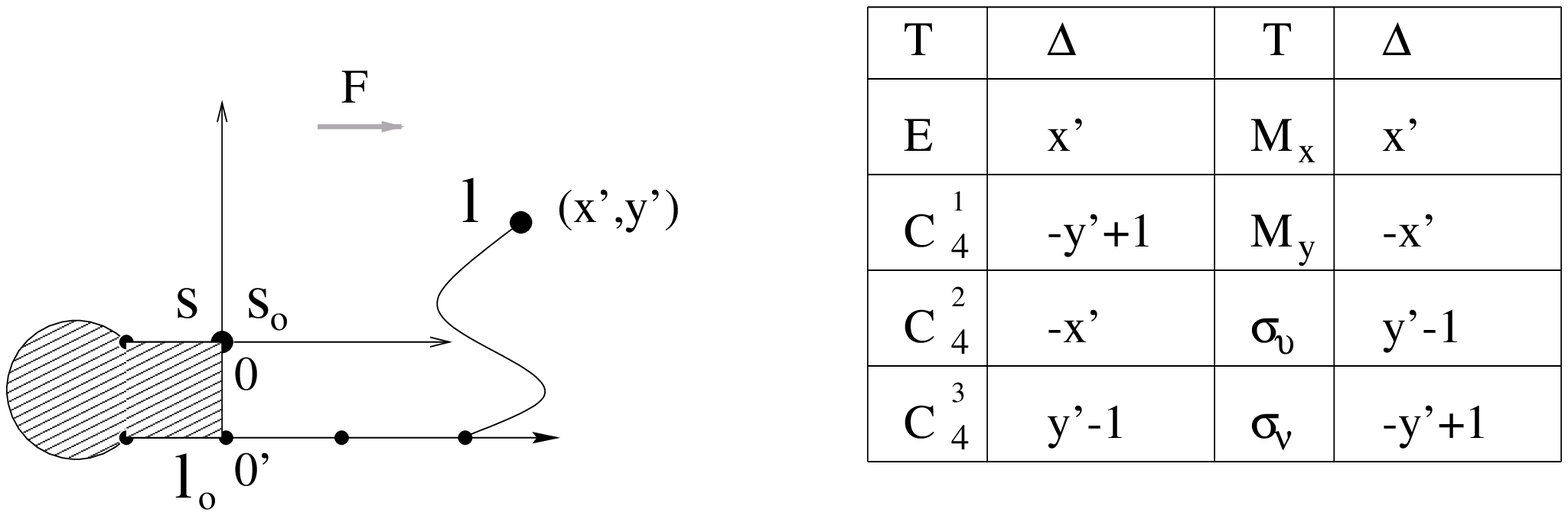}
\caption{One class of conformations for the 2nd tail $(l_o,l)$ and 
CG $(s_0,l_o)$ containing outmost type 1 link, where the tail type 
is stiff and extends upward \protect\cite{chen98}. The force stretching 
end $l$ located at $(x',y')$ on frame $O'$ is translated to point $(x',y'-1)$ 
on frame $O$, and the corresponding EED $\Delta$ is $x'$. Then we distribute 
the conformation to lattice plane by eight square symmetry transformations 
(see Fig.~\ref{hairpinforc}). These EEDs are tabulated respectively, where 
the character ``T" represents transformation elements. 
}
\label{complextail}
\end{center}
\end{figure}

For hairpin conformations, the conformations of tails are determined by 
two factors: the outermost link type and the length of CG. To 
account for excluded interactions, the tails have been classified into two 
types, stiff (s) and flex (f)\cite{chen98}. 
Correspondingly, the number of conformations is $n^{t_i}(N_i)$, where $t_i$ 
(s or f) are the type of $i$th (1st or 2nd) tail with length $N_i$, or 
$N_i$-step OSAW. In order to calculate EEDs of whole chain , we introduce 
additional definitions: $n_x^{t_i}(N_i;m)$, the number of conformations 
for $i$th OSAW of type $t_i$ whose final $x$ coordinates are $m$; and 
$n_y^{t_i}(N_i;m)$ is similar except $y$ coordinate. It is ease to find  
$n^{t_i}(N_i)=\sum_m n_x^{t_i}(N_i;m)=\sum_m n_y^{t_i}(N_i;m)$. Here we 
illustrate the calculation of 
diagonal matrix element ${\bf \Pi}(l,l_0;s_0,s)_{11}$ 
when the outmost link is of type $1$.
\subsection{$s=s_0$, $l\ne l_0$}
In this case, the length of the 2nd tail is $N_2=l-l_o$. There are six classes 
of possible conformations for CG-tail complexes\cite{chen98}. In 
Fig.~\ref{complextail}, we show one of them. Since the length of CG also 
affects exclude volume interaction, we distinguish different CG sizes. 

(1) $l_o-s_o=3$.
In this case, all six classes of conformations are viable. Therefore, 
\begin{eqnarray}
\Pi\left[s,s_0;l_0,l|f\right]_{11}&=&2\times2\left[  \sum\limits_y 
n_y^{f_2}(N_2;y)\left( \cosh\beta f y + \cosh\beta f(y+1)\right)\right.
\nonumber\\
&&+\sum\limits_x n_x^{f_2}(N_2;x)\left( \cosh\beta f x+\cosh\beta f(x+1)
\right)\nonumber\\
&&+\sum\limits_y n_y^{s_2}(N_2;y)\left( 2\cosh\beta f y+\cosh\beta f(y+1)+
\cosh\beta f(y-1)\right)\nonumber\\
&&+\sum\limits_x n_x^{s_2}(N_2;x)\left( 2\cosh\beta f x+2\cosh\beta f(x+1)
\right)\nonumber\\ 
&&\left.-\left(1+\cosh\beta f+ \cosh\beta f N_2 + \cosh\beta f(N_2+1)\right)
\right], 
\end{eqnarray}
where one of coefficients $2$ is {\it degeneracy degree} along force direction, 
and another is from combining exponentials to hyperbolic cosine function, 
the negative part is to eliminate the straight tail conformations, which 
are counted repeatedly.

(2) $l_o-s_o>3$.
There are five classes of conformations are involved. We have
\begin{eqnarray}
\Pi\left[s,s_0;l_0,l|f\right]_{11}&=&2\times2\left[  \sum\limits_y 
n_y^{f_2}(N_2;y)\left( \cosh\beta f y + \cosh\beta f(y+1)\right)\right.
\nonumber\\
&&+\sum\limits_x n_x^{f_2}(N_2;x)\left( \cosh\beta f x+\cosh\beta f(x+1)
\right)\nonumber\\
&&+\sum\limits_y n_y^{s_2}(N_2;y)\left( \cosh\beta f y+\cosh\beta f(y+1)+
\cosh\beta f(y-1)\right)\nonumber\\
&&+\sum\limits_x n_x^{s_2}(N_2;x)\left( 2\cosh\beta f x+\cosh\beta f(x+1)
\right)\nonumber\\ 
&&\left.-\left(\cosh\beta f+ \cosh\beta f N_2 \right)
\right].
\end{eqnarray}

\subsection{$s\ne s_0$, $l\ne l_0$}
A large number of CG-tail complexes have been listed in Ref. \cite{chen98}. 
Since there 
are two tails, we define $N_1=s_o-s$ and $N_2=l-l_o$. According the length of 
CG part, we then have following results.

(1) $l_o-s_o=3$.
This case is more cumbersome than above cases. We divide 
$\left.\Pi\left[s,s_0;l_0,l\right|f\right]_{11}$ into the sum over $\Pi^{(j)}$, 
$j=1,\cdots,5$ such as 
\begin{eqnarray}
\left.\Pi\left[s,s_0;l_0,l\right|f\right]_{11}=2\times2\sum_{j=1}^5 \Pi^{(j)}, 
\end{eqnarray}
where 
\begin{eqnarray}
\Pi^{(1)}&=&\sum\limits_{y_1} 
\sum\limits_{y_2} n_y^{s_1}(N_1;y_1)n_y^{s_2}(N_2;y_2)
\left( 2\cosh\beta f(y_1+y_2) + \cosh\beta f(y_1-y_2)\right)\nonumber\\
&&+\sum\limits_{y_1} \sum\limits_{x_2} 
n_y^{s_1}(N_1;y_1)n_x^{s_2}(N_2;x_2)\left( 2\cosh\beta f(y_1+x_2+1) + 
\cosh\beta f(y_1-x_2-1)\right.\nonumber \\
&&+\left.\cosh\beta f(y_1+x_2)+\cosh\beta f(y_1-x_2)\right)\nonumber\\
&&+\sum\limits_{x_1} \sum\limits_{y_2}n_x^{s_1}(N_1;x_1)n_y^{s_2}(N_2;y_2)
\left( 2\cosh\beta f(x_1+y_2+1) + \cosh\beta f(x_1-y_2+1)\right.\nonumber\\
&&+\left.2\cosh\beta f(x_1+y_2)+\cosh\beta f(x_1-y_2)\right)\nonumber\\
&&+\sum\limits_{x_1} \sum\limits_{x_2}n_x^{s_1}(N_1;x_1)n_x^{s_2}(N_2;x_2)
\left( 3\cosh\beta f(x_1+x_2+1) \right),
\end{eqnarray}
\begin{eqnarray}
\Pi^{(2)}&=&\sum\limits_{y_1} \sum\limits_{y_2}n_y^{s_1}(N_1;y_1)n_y^{f_2}(N_2;y_2)
\left( \cosh\beta f(y_1+y_2)\right)\nonumber\\
&&+\sum\limits_{y_1} \sum\limits_{x_2}n_y^{s_1}(N_1;y_1)n_x^{f_2}(N_2;x_2)
\left( \cosh\beta f(y_1+x_2+1) + \cosh\beta f(y_1+x_2)\right.\nonumber\\
&&+\left.\cosh\beta f(y_1-x_2)\right)\nonumber\\
&&+\sum\limits_{x_1} \sum\limits_{y_2}n_x^{s_1}(N_1;x_1)n_y^{f_2}(N_2;y_2)
\left( 2\cosh\beta f(x_1+y_2+1) + \cosh\beta f(x_1-y_2)\right)\nonumber\\
&&+\sum\limits_{x_1} \sum\limits_{x_2}n_x^{s_1}(N_1;x_1)n_x^{f_2}(N_2;x_2)
\left( \cosh\beta f(y_1+y_2)\right),
\end{eqnarray}
\begin{eqnarray}
\Pi^{(3)}&=&\sum\limits_{y_1} \sum\limits_{y_2}n_y^{f_1}(N_1;y_1)n_y^{s_2}(N_2;y_2)
\left( \cosh\beta f(y_1+y_2)\right)\nonumber\\
&&+\sum\limits_{y_1} \sum\limits_{x_2}n_y^{f_1}(N_1;y_1)n_x^{s_2}(N_2;x_2)
\left( 2\cosh\beta f(y_1+x_2+1) + \cosh\beta f(y_1-x_2)\right)\nonumber\\
&&+\sum\limits_{x_1} \sum\limits_{y_2}n_x^{f_1}(N_1;x_1)n_y^{s_2}(N_2;y_2)
\left( \cosh\beta f(x_1+y_2+1) + \cosh\beta f(x_1+y_2)\right.\nonumber\\
&&+\left.\cosh\beta f(x_1-y_2)\right)\nonumber\\
&&+\sum\limits_{x_1} \sum\limits_{x_2}n_x^{f_1}(N_1;x_1)n_x^{s_2}(N_2;x_2)
\left( \cosh\beta f(x_1+x_2+1)\right),
\end{eqnarray}
\begin{eqnarray}
\Pi^{(4)}&=&\sum\limits_{y_1} \sum\limits_{x_2}n_y^{f_1}(N_1;y_1)n_x^{f_2}(N_2;x_2)
\left( \cosh\beta f(y_1+x_2+1)+\cosh\beta f(y_1-x_2)\right)\nonumber\\
&&+\sum\limits_{x_1} \sum\limits_{y_2}n_x^{f_1}(N_1;x_1)n_y^{f_2}(N_2;y_2)
\left( \cosh\beta f(x_1+y_2+1)+\cosh\beta f(x_1-y_2)\right),
\end{eqnarray}
\begin{eqnarray}
\Pi^{(5)}&=&-\left[
\sum\limits_{y_1}n_y^{s_1}(N_1;y_1)\left(\cosh \beta f(y_1+N_2+1)+
\cosh \beta f(y_1+N_2)+ \cosh \beta fy_1\right)\right. \nonumber\\
&&+\sum\limits_{x_1}n_x^{s_1}(N_1;x_1)\left(\cosh \beta f(x_1+N_2+1)+
\cosh \beta f(x_1+1)+ \cosh \beta fx_1\right)\nonumber\\
&&+\sum\limits_{y_2}n_y^{s_2}(N_2;y_2)\left(\cosh \beta f(y_2+N_1+1)+
\cosh \beta f(y_2+N_1)+ \cosh \beta fy_2\right)\nonumber\\
&&+\sum\limits_{x_2}n_x^{s_2}(N_2;x_2)\left(\cosh \beta f(x_2+N_1+1)+
\cosh \beta f(x_2+1)+ \cosh \beta fx_2\right)\nonumber\\
&&+\sum\limits_{y_1}n_y^{f_1}(N_1;y_1)\left(\cosh \beta f(y_1+N_2+1)
\right)
+\sum\limits_{x_1}n_x^{f_1}(N_1;x_1)\left(\cosh \beta fx_1\right)
\nonumber\\
&&\left.+\sum\limits_{y_2}n_y^{f_2}(N_2;y_2)\left(\cosh \beta f(y_2+N_1+1)
\right)+
\sum\limits_{x_2}n_x^{f_2}(N_1;x_2)\left(\cosh \beta fx_2\right)
\right]. 
\end{eqnarray}

(2) $l_o-s_o>3$.
\begin{eqnarray}
\Pi\left[s,s_0;l_0,l|f\right]_{11}&=&2\times\left[
\sum\limits_{y_1} \sum\limits_{x_2}n_y^{s_1}(N_1;y_1)n_x^{s_2}(N_2;x_2)
\left( \cosh\beta f(y_1+x_2+1)+ \cosh\beta f(y_1-x_2+1)\right)\right.\nonumber\\
&&+\sum\limits_{x_1} \sum\limits_{y_2}n_x^{s_1}(N_1;x_1)n_y^{s_2}(N_2;y_2)
\left( \cosh\beta f(x_1-y_2) + \cosh\beta f(x_1+y_2+1)\right)\nonumber\\
&&+\sum\limits_{y_1} \sum\limits_{x_2}n_y^{s_1}(N_1;y_1)n_x^{f_2}(N_2;x_2)
\left( \cosh\beta f(y_1+x_2+1) + \cosh\beta f(y_1-x_2)\right)\nonumber\\
&&+\sum\limits_{x_1} \sum\limits_{y_2}n_x^{s_1}(N_1;x_1)n_y^{f_2}(N_2;y_2)
\left( \cosh\beta f(x_1-y_2) + \cosh\beta f(x_1+y_2+1)\right)\nonumber\\
&&+\sum\limits_{y_1} \sum\limits_{x_2}n_y^{f_1}(N_1;y_1)n_x^{s_2}(N_2;x_2)
\left( \cosh\beta f(y_1-x_2) + \cosh\beta f(y_1+x_2+1)\right)\nonumber\\
&&+\sum\limits_{x_1} \sum\limits_{y_2}n_x^{f_1}(N_1;x_1)n_y^{s_2}(N_2;y_2)
\left( \cosh\beta f(x_1+y_2+1) + \cosh\beta f(x_1-y_2)\right)\nonumber\\
&&+\sum\limits_{x_1} \sum\limits_{y_2}n_x^{f_1}(N_1;x_1)n_y^{f_2}(N_2;y_2)
\left( \cosh\beta f(x_1-y_2) + \cosh\beta f(x_1+y_2+1)\right)\nonumber\\
&&+\sum\limits_{y_1} \sum\limits_{x_2}n_y^{f_1}(N_1;y_1)n_x^{f_2}(N_2;x_2)
\left.\left( \cosh\beta f(y_1+x_2+1) + \cosh\beta f(y_1-x_2)\right)\right].
\end{eqnarray}
These equations are so boring that we are not ready to list all 
$\Pi[s,s_o;l_o,l|f]$ (added up to 15 cases). If one sets force vanishing, 
all equations will return to Chen and Dill results for hairpin 
case\cite{chen98}. 

\section{extrapolation technique for one dimension end-to-end distance
distribution of N-step open self-avoiding walk}
\label{extrapolate}
The knowledge about the EED of N-step OSAW is important for our studies, 
especially, when N is larger enough, the exact enumeration can 
not bear. Fortunately, similar issues has been discussed 
early, where an asymptotic formula about the distribution of SAW was 
fitted\cite{domb,fisher66}, In this paper, we follows the similar 
approach to find the distribution density of the OSAW.

\begin{figure}[htpb]
\begin{center}
\includegraphics[width=0.4\columnwidth]{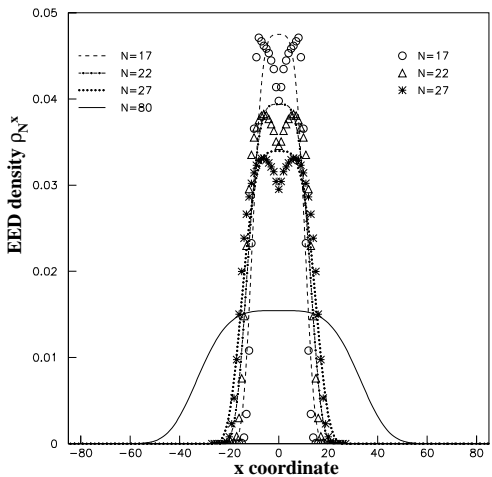}
\caption{Comparing extrapolation formula Eq.~\protect\ref{density} 
(lines) with exact enumeration results (symbols) about EED density of OSAW. 
The solid line is of 80-step walk.}
\label{extrapolation}
\end{center}
\end{figure}

Let $C_N$ be the total number of N-steps OSAWs, and 
let $C_x(N;m)$ be the number of walks which reach $m$ at x coordinate. 
Then EED density is defined as $\rho_N^x(m)=C_x(N;m)/C_N$. According to the 
density form suggested by Domb {\it et al.}, $\rho_N^x(m)$ is written as 
\begin{eqnarray}
\rho_N^x(m)=\frac{0.32070}{<x_N^2>^{1/2}}\exp(-\frac{0.11423 x^4}{<x_N^2>^2})
\label{density}
\end{eqnarray}
where $<x_N^2>$ is one dimensional mean-square EED. The mean square 
$<x_N^2>$ can be solved by extrapolation technique\cite{fisher61}. Here we 
only give our numerical result, $<x_N^2>=0.75 N^{1.45}$. To computer 
$C_x(N;m)$, we need $C_N$ from the definition of $\rho_N^x$, whose asymptotic 
formula has been provided by Chen and Dill\cite{chen98}. Several examples 
about EED density of OSAW are shown in Fig.~\ref{extrapolation}

We note that there is a dip in the value of enumeration at the origin 
arising from the restriction of no returns to the origin. It might be 
possible to provide more refined approximation\cite{chay}, but in view of the 
general errors involved in the extrapolation process we have not 
considered this worth pursuing\cite{domb}. In our studies, because chain 
molecule would like to extend under the force, the approximation 
is also reasonable.

\end{document}